 \documentclass[preprint2]{aastex}

\usepackage{longtable}

\shorttitle{Chemical clues on the formation of planetary systems}
\shortauthors{Delgado Mena  et al.}

\begin{document}


\title{Chemical clues on the formation of planetary systems \\
\begin{normalsize}C/O vs Mg/Si for HARPS GTO sample\end{normalsize}}


\author{E. Delgado Mena\altaffilmark{1.2}, G.
Israelian\altaffilmark{1,2}, J.~I. Gonz\'alez
Hern\'andez\altaffilmark{1,2}, J.~C. Bond\altaffilmark{3}, N.~C.
Santos\altaffilmark{4,5,6}, S. Udry\altaffilmark{6} and M.
Mayor\altaffilmark{6}} 





\altaffiltext{1}{Instituto de Astrof\'isica de Canarias, 38200 La Laguna, Tenerife, Spain: edm@iac.es}
\altaffiltext{2}{Departamento de Astrof\'isica, Universidad de La Laguna, 38205 La Laguna, Tenerife, Spain}
\altaffiltext{3}{Planetary Science Institute, 1700 E. Fort Lowell, Tucson, AZ 85719, USA}
\altaffiltext{4}{Centro de Astrof\'isica, Universidade do Porto, Rua das Estrelas, 4150-762 Porto, Portugal}
\altaffiltext{5}{Departamento de F\'isica e Astronomia, Faculdade de Ci\^encias, Universidade do Porto, Portugal}
\altaffiltext{6}{Observatoire de Gen\`eve, Universit\'e de Gen\`eve, 51 Ch. des Mailletes, 1290 Sauverny, Switzerland}

\begin{abstract}
Theoretical studies suggest that C/O and Mg/Si are the most important
elemental ratios in determining the mineralogy of terrestrial planets.
The C/O ratio controls the distribution of Si among carbide and oxide
species, while Mg/Si gives information about the silicate mineralogy.
We present a detailed and uniform study of C, O, Mg and Si abundances
for 61 stars with detected planets and 270 stars without detected planets
from the homogeneous high-quality unbiased HARPS GTO sample, together with 39
more planet-host stars from other surveys. We determine 
these important mineralogical ratios and investigate the nature of 
the possible terrestrial planets that could have formed in those 
planetary systems. 
We find mineralogical ratios quite different from those of the Sun, showing that
there is a wide variety of planetary systems which are not similar to
Solar System. Many of planetary host stars present a Mg/Si
value lower than 1, so their planets will have a high Si content to
form species such as MgSiO$_{3}$. This type of composition can have
important implications for planetary processes like plate tectonics,
atmospheric composition or volcanism. 
\end{abstract}





\section{Introduction}

The study of extrasolar planets has been a new
exciting field of astrophysics for ten years now. More than 450 planets are known in 385
planetary systems. In addition, more than 80 planets out of these 450
transit their host stars and in the last few years more than 40
planets with minimum masses between 2 and 20 $M_{\earth}$ have
been discovered. The study of the photospheric stellar abundances of 
their parent stars is the key to understand how and which of the
protoplanetary clouds form planets and which do not. These studies
also help us to investigate the internal and atmospheric structure and
composition of extrasolar planets.\\ 

\begin{figure*}[ht]
\centering
\includegraphics[width=7.8cm]{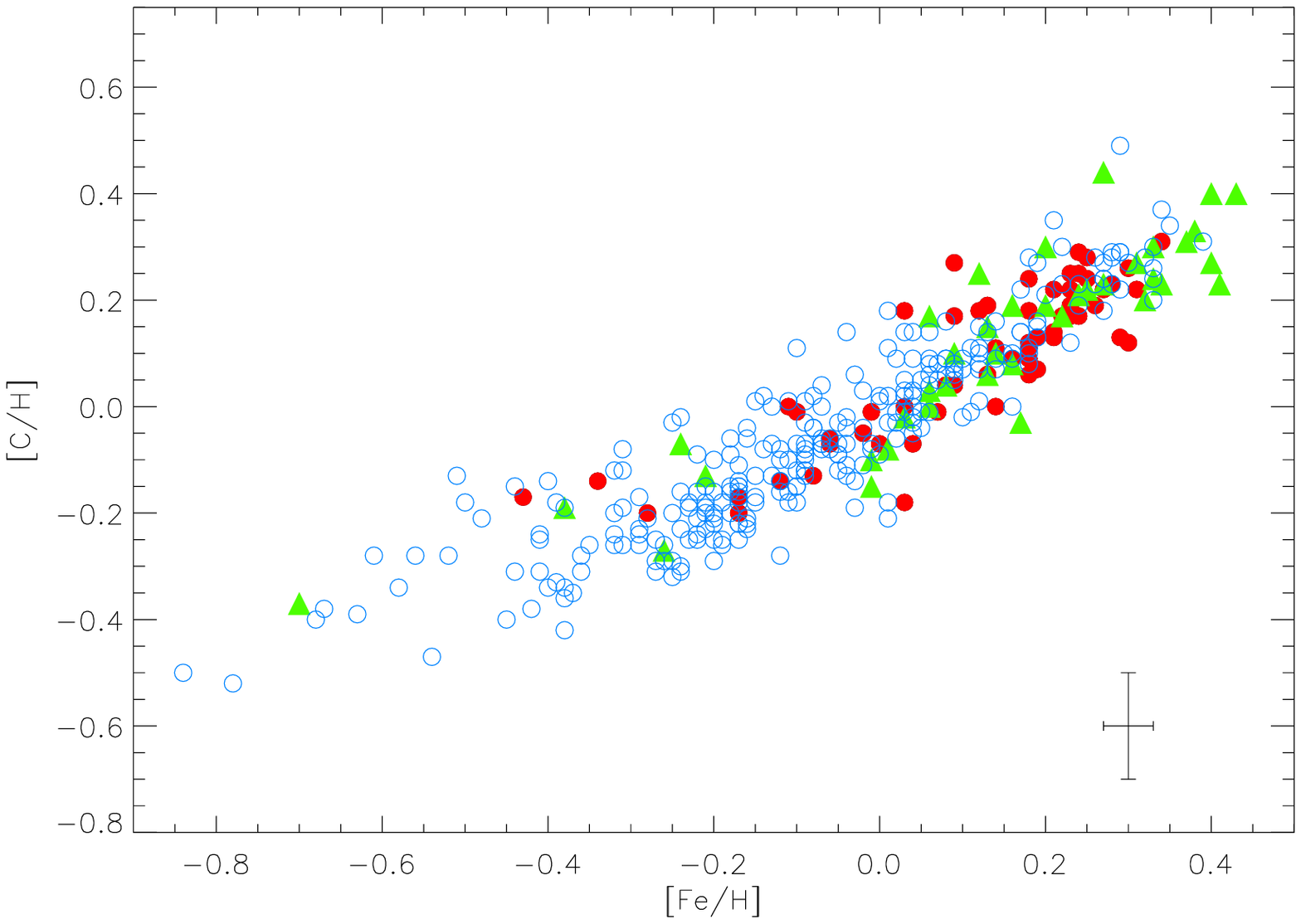}
\includegraphics[width=7.8cm]{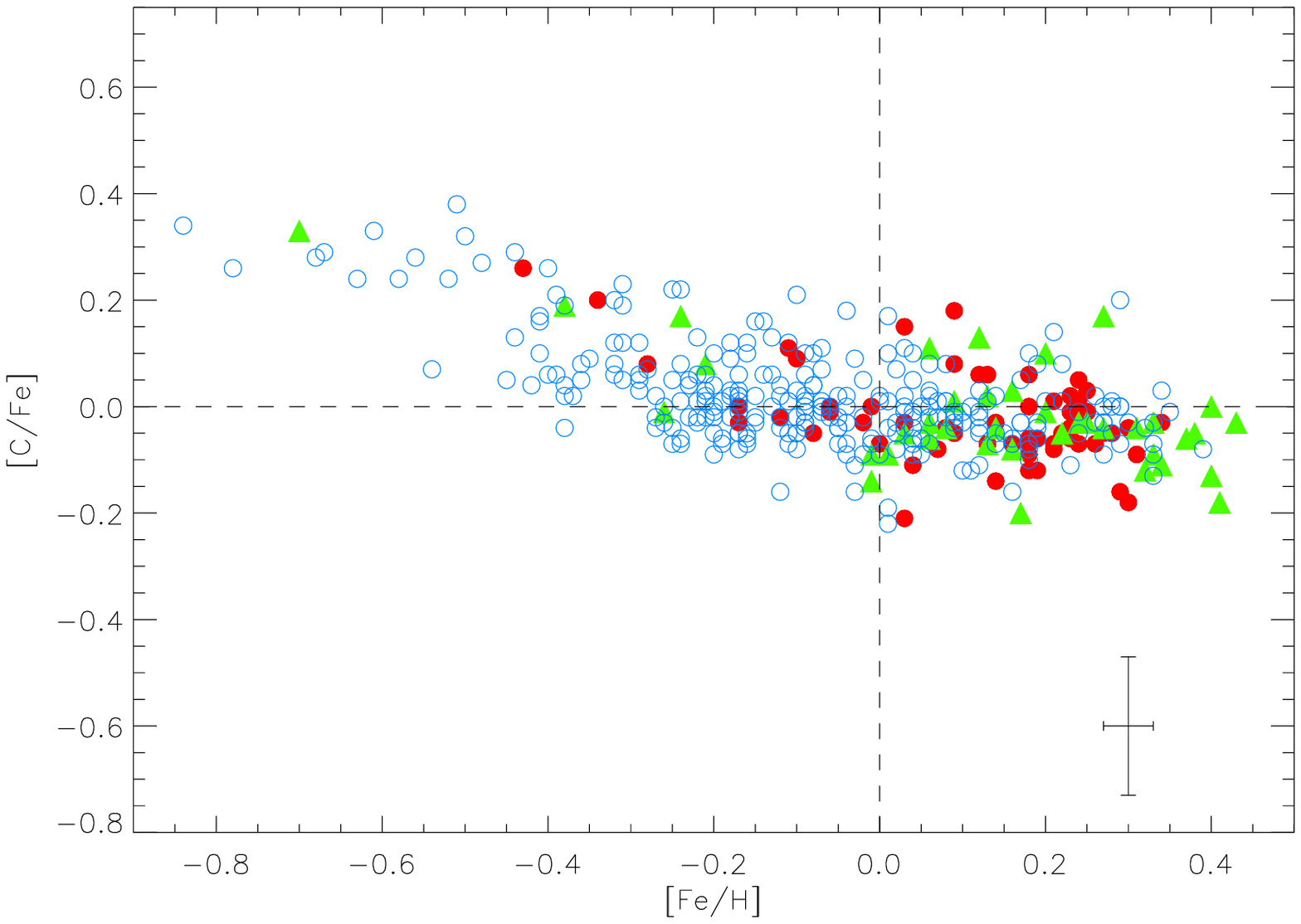}
\caption{[C/H] vs [Fe/H] and [C/Fe] vs [Fe/H] for stars with (red filled circles) and without (blue open circles) detected planets from the HARPS GTO sample. Green triangles are stars with planets from other surveys.}
\label{carb}
\end{figure*}

One remarkable characteristic of planet host stars is that they are
considerably metal rich when compared with single field dwarfs
\citep{Gonzalez98, Santos00, Gonzalez01, Santos01, santos04, Fischer}. Two main
explanations have been suggested to clarify this difference. The
first of these is that the origin of this metallicity excess is primordial, so the
more metals you have in the proto-planetary disk, the higher should be
the probability of forming a planet. On the other hand, this excess
might be produced by accretion of rocky material by the star some time
after it reached the main-sequence. If pollution were the responsible
for the enhanced metallicity of planet hosts, we would expect to find
higher metallicities as the convective envelope mass decreases, but
no such trend has been found. In addition, transit detections have shown
that the mass of heavy elements in the planets appears to be
correlated with the metallicity of their parent stars \citep{guillot}.
A recent work by \citet{Mordasini} finds that distributions of
planetary systems are well reproduced using core-accretion models,
which are dependent on dust content of the disk, thus supporting the
primordial origin of supersolar metallicity in stars with planets.
Recent studies on chemical abundances in stars with and without planets showed no important differences in [X/Fe] vs. [Fe/H] trends between both groups of stars \citep{Takeda07,bond08,neves,jonay}. However, other works have reported less statistically significant enrichments in other species such as C, Na, Si, Ni, Ti, V, Co, Mg and Al \citep{Gonzalez01,Santos00,Sadakane,Bodaghee,Fischer,Beirao,gilli,bond06,Gonzalez07} or even important enrichments in Si and Ni \citep{Robinson}.\\

These results have important implications for models of giant
planet formation and evolution. There are two major planet formation
models: the core accretion model \citep{Pollack}, more likely to
form planets in the inner disk, and the disk instability model
\citep{Boss}, which is in better agreement with the conditions in the
extended disk. In the first model, planets are formed by the
collisional acumulation of planetesimals by a growing solid core,
followed by accretion of a gaseous envelope onto the core. In the
second scenario, a gravitationally unstable region in a protoplanetary disk
forms self-gravitating clumps of gas and dust, within which the dust
grains coagulate and sediment to form a central core \citep{Boss}. In
the core accretion model, planet formation is dependent on the dust
content of the disk \citep{Pollack} while in the disk instability
model it is not \citep{Boss02}. Present observations are thus more
compatible with core accretion model although they do not exclude disk
instability.\\ 

Theoretical studies suggest that C/O and Mg/Si are the most important
elemental ratios in determining the mineralogy of terrestrial planets
and they can give us information about the composition of these
planets. The C/O ratio controls the distribution of Si among carbide
and oxide species, while Mg/Si gives information about the silicate
mineralogy \citep{bond_earth, bond_sim}. \citet{bond_sim} carried out
simulations of planet formation where the chemical composition of the
protoplanetary cloud was taken as an input parameter. Terrestrial
planets were found to form in all the simulations with a wide variety
of chemical compositions so these planets might be very different from
the Earth. In this paper we will present C/O and Mg/Si ratios in a
sample of stars with and without detected planets using new 
high quality spectra in order to investigate the mineralogical
characteristics of those systems.\\ 

\section{Observations}

The HARPS GTO sample is composed of 451 FGK stars selected from a
volume-limited stellar sample observed by the CORALIE spectrograph at
La Silla observatory. These stars are slowly-rotating, non-evolved,
and low-activity stars that presented no obvious radial-velocity
variations at the level of the CORALIE measurement precision. For more
details we point the reader to a description of the sample by
\citet{mayor}. This sample is composed of high resolution, 
high signal-to-noise spectra for 71 stars with planets and 380 with no known giant
planets with effective temperatures from 4500 K to 6500 K. Precise stellar parameters were taken from \citet{sousa08},
with uncertainties of the order of 30 K for T$_{\rm eff}$, 0.06 dex for
log \textit{g}, 0.08 km s$^{-1}$ for $\xi_{t}$ and 0.03 dex for
[Fe/H]. To improve the statistics we added high quality spectroscopic
observations for 42 stars hosting planets from the CORALIE survey, 
using the same spectral tools to determine their stellar parameters
\citep{santos04, santos05}, and thus ensuring that the final sample is
homogeneous. 

  \begin{figure}[ht]
   \centering
   \includegraphics[width=6.cm]{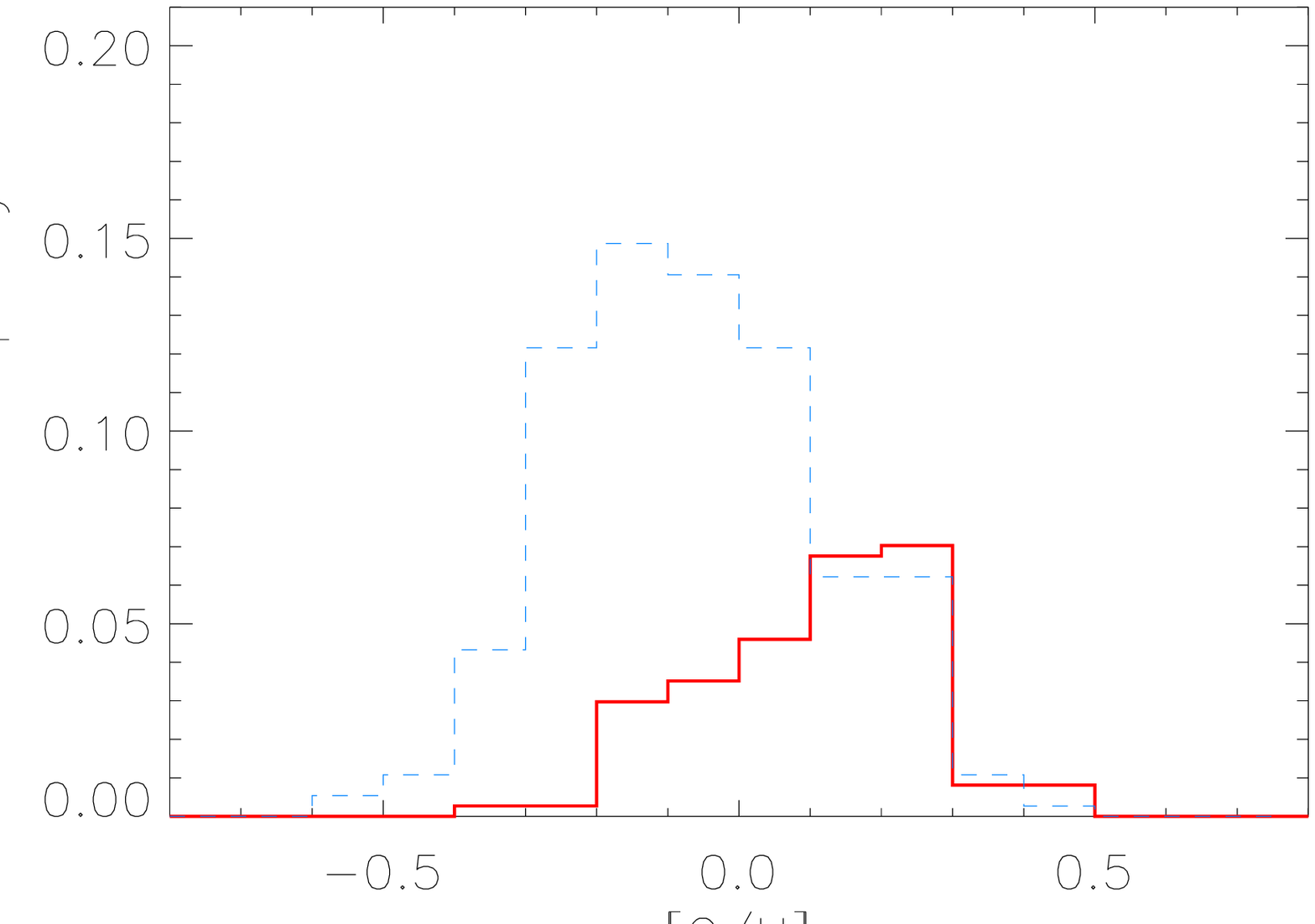}
   \includegraphics[width=6.cm]{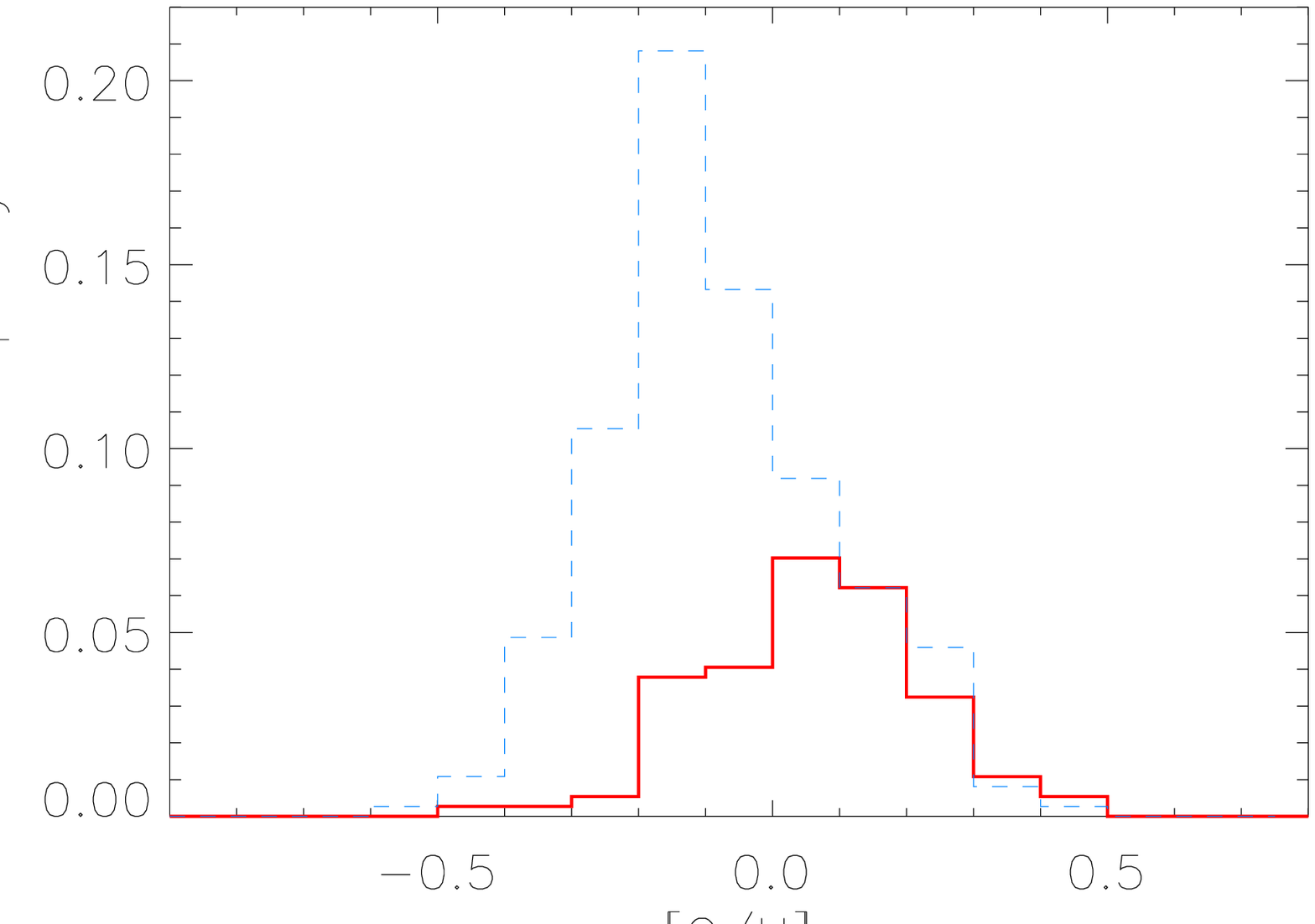}
      \caption{[C/H] and [O/H] distributions for stars with (red line) and without (blue dashed line) detected planets.}
         \label{histograma}
   \end{figure}

\section{Analysis}

For all the elements we performed a standard LTE analysis with 
the 2002 revised version of the spectral synthesis code MOOG 
\citep{sneden} and a grid of Kurucz ATLAS9 atmospheres with 
overshooting \citep{kur93}, by measuring the equivalent width 
(EW) of the different lines with the ARES 
program\footnote{The ARES code can be downloaded at
http://www.astro.up.pt/~sousasag/ares/} \citep{sousa_ares}. All the abundances are listed in Tables~\ref{lista_planetas1},~\ref{lista_planetas2},~\ref{lista_alex},~\ref{lista_comp1},~\ref{lista_comp2},~\ref{lista_comp3},~\ref{lista_comp4},~\ref{lista_comp5} and ~\ref{lista_comp6}.

\begin{figure}[h]
   \centering
   \includegraphics[width=8cm]{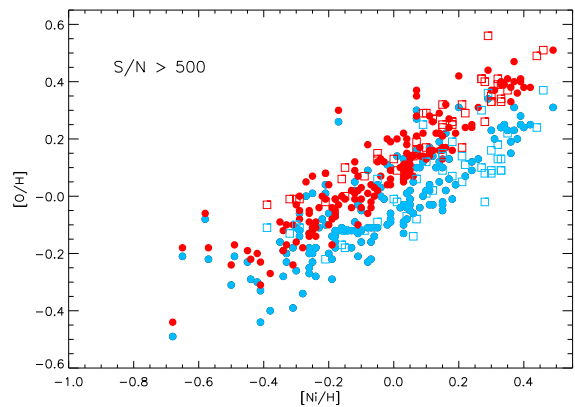}
      \caption{[O/H] vs [Ni/H] for stars with (squares) and without (circles) detected planets from the HARPS GTO sample. Red and blue symbols correspond to O abundance with and without the contribution of Ni, respectively.}
         \label{niquel}
   \end{figure}
\begin{figure*}[ht]
   \centering
   \includegraphics[width=7.8cm]{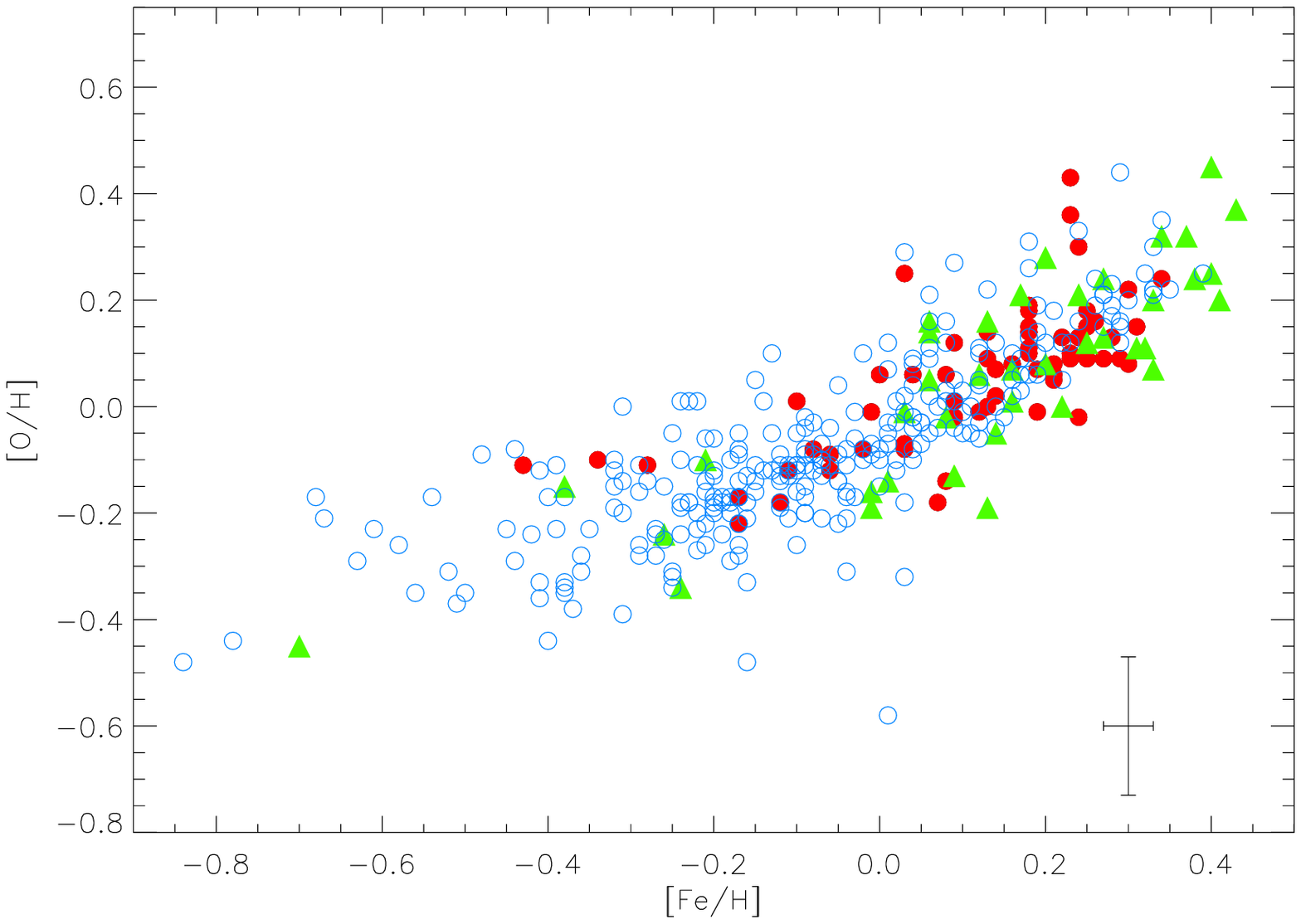}
   \includegraphics[width=7.8cm]{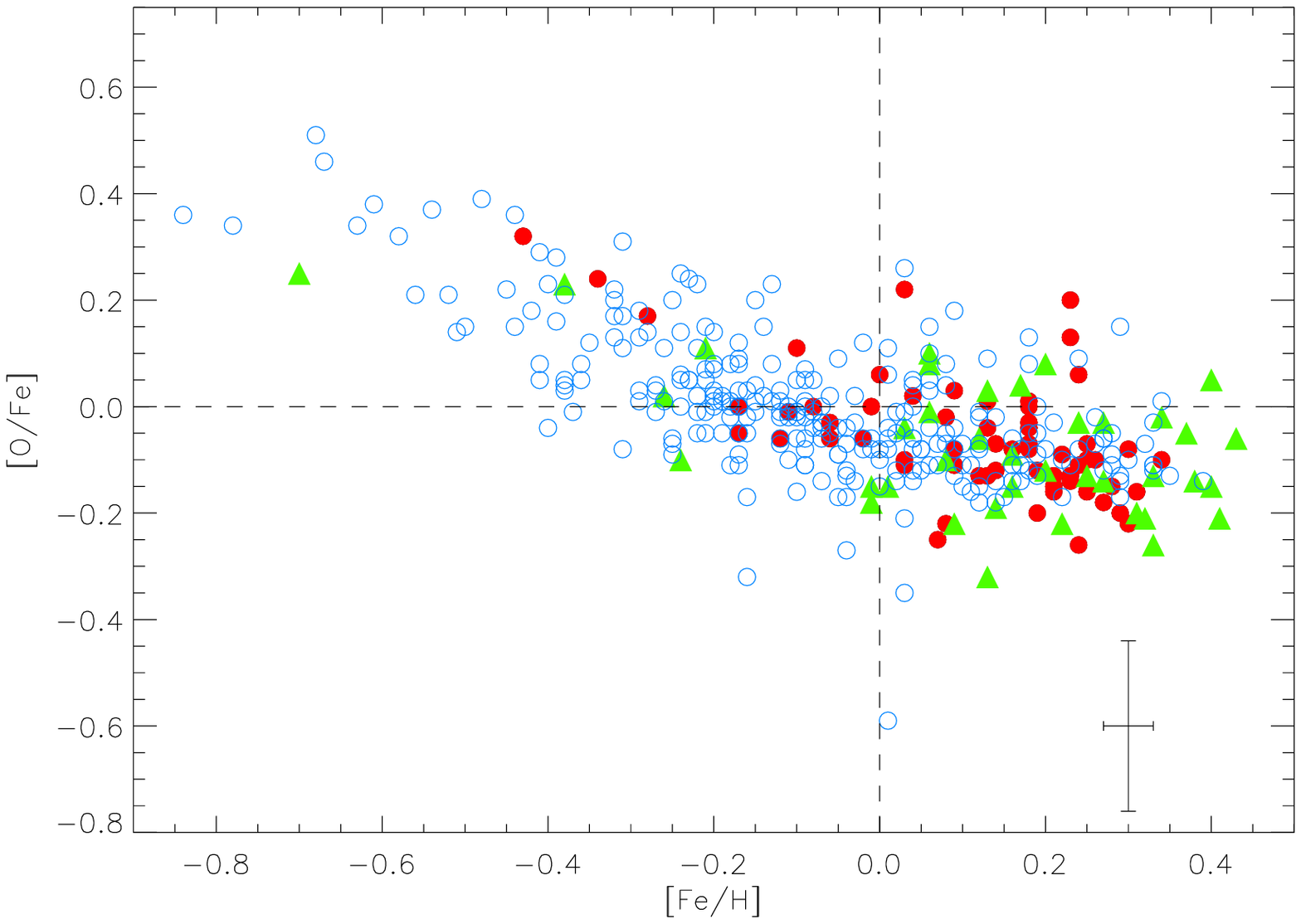}
      \caption{[O/H] vs [Fe/H] and [O/Fe] vs [Fe/H] for stars with (red filled circles) and without (blue open circles) detected planets from the HARPS GTO sample. Green triangles are stars with planets from other surveys.}
         \label{oxig}
   \end{figure*}
\begin{figure*}[ht]
   \centering
   \includegraphics[width=7.8cm]{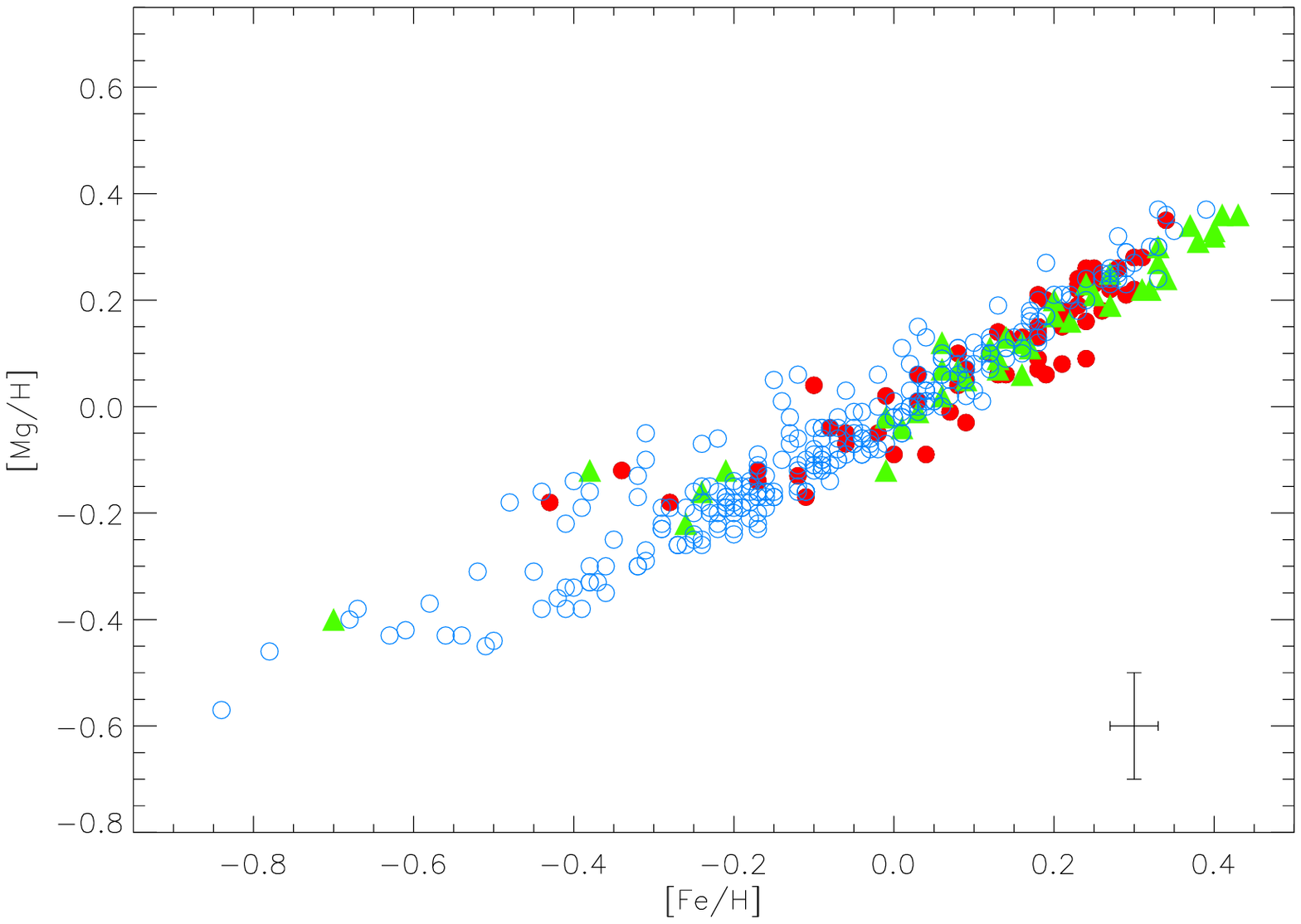}
   \includegraphics[width=7.8cm]{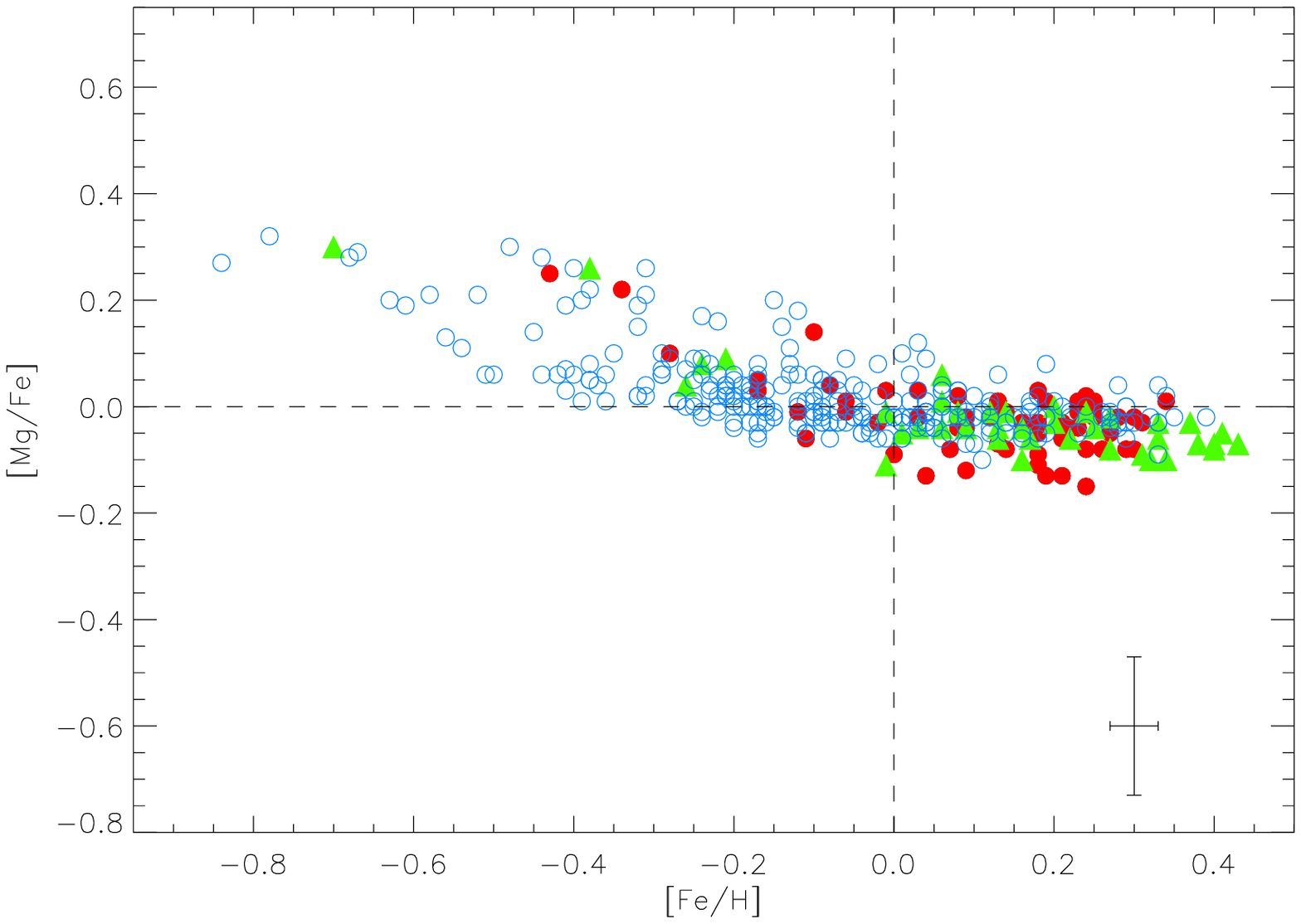}
      \caption{[Mg/H] vs [Fe/H] and [Mg/Fe] vs [Fe/H] for stars with (red filled circles) and without (blue open circles) detected planets from the HARPS GTO sample. Green triangles are stars with planets from other surveys.}
         \label{magn}
   \end{figure*}

In Figs.~\ref{carb},~\ref{oxig},~\ref{niquel},~\ref{magn} 
and~\ref{sil} we display at the left-bottom corner
of each panel the average error bars for the element abundances,
[X/H], or abundances ratios [X/Fe].

The errors in the element abundances, [X/H], show
their sensitivity to the uncertainties in the effective temperature
($\Delta_{T_{\mathrm{eff}}}$), surface gravity ($\Delta_{\log g}$),
microturbulence ($\Delta_{\xi}$), continuum placement and the dispersion of the
measurements from different spectral features ($\Delta_{\sigma}$). 
The errors $\Delta_{\sigma}$ were estimated as $\Delta_{\sigma}
=\sigma/\sqrt{N}$, where $\sigma$ is the standard deviation of the 
$N$ measurements. We estimate the total error by adding in quadrature 
all these uncertainties. 

The errors in the abundance ratios, [X/Fe], were determined taking
into account the differences between the sensitivities 
of the resulting abundances to changes in assumed atmospheric
parameters and the dispersion of the abundances from individual lines
of each element. 

 \begin{center}
\begin{table}[ht]
\caption{Atomic parameters for lines of  \ion{C}{1}, [\ion{O}{1}] and \ion{Ni}{1}.}
\label{lineas}
\centering
\begin{tabular}{cccc}
\hline
\noalign{\medskip} 
Element & $\lambda$ (\AA{}) & $\chi_{l}$ (eV) & log \textit{gf}\\
\noalign{\medskip} 
\hline
\hline
\noalign{\medskip} 
\ion{C}{1}   & 5052.160 & 7.68 & -1.420 \\
\ion{C}{1}   & 5380.340 & 7.68 & -1.710 \\
$[$\ion{O}{1}$]$ & 6300.230 & 0.00 & -9.689 \\
\ion{Ni}{1}  & 6300.399 & 4.27 & -2.310 \\
\hline
\end{tabular}
\end{table}
\end{center}

\subsection{Carbon\label{seccarbon}}

To obtain carbon abundances we used two unbleded lines at $\lambda$
5380.3 \AA{} and $\lambda$ 5052.2 \AA{}. For the coolest stars, 5052.2
\AA{} line becomes very weak and the abundance is calculated using only
5380.3 \AA{} line is very high, so we removed from the samples all
stars with T$_{\rm eff}$ $<$ 5100 K. The wavelengths and excitation
energies of the lower levels were taken from VALD database
\citep{vald}. The oscillator strengths, $\log gf$ values, were 
adjusted using the EWs obtained from the Kurucz Solar Atlas and a 
solar model with $T_{\rm eff} = 5777$~K, $\log g = 4.44$ and 
$\xi_{t} = 1$~km~s$^{-1}$ to get $\log \epsilon({\rm
C})_{\sun}=8.56$\footnote{log 
$\epsilon$(X) = log[(N(X)/N(H)] + 12} \citep{anders}, which is the
solar value used for the differential analysis (see Table
\ref{lineas}). We also calculated solar C abundance using a solar
Harps spectrum\footnote{The HARPS solar spectra can be downloaded
at
http://www.eso.org/sci/facilities/lasilla/instruments/harps/\\/inst/monitoring/sun.html}
(daytime sky spectrum) and the same model, obtaining 
$\log \epsilon({\rm C})_{\sun}= 8.52$. 
We note here that the spectral lines in solar spectra obtained on the
daytime sky are known to exhibit EW and line depth changes \citep[e.g.][]{gray00}.
This may explain these different C abundances. We may 
refer to the work by \citet{jonay} to see the
differences in element abundances from slightly different solar HARPS spectra
and those of the solar ATLAS spectrum. 
In this work we will use ATLAS solar values as reference values.
However, we will plot both solar values in the C/O vs Mg/Si figure (see Figs. \ref{planetas} and \ref{CO}).\\ 

In Fig. \ref{carb} we show the [C/H] and [C/Fe] ratios as a function
of [Fe/H]. The samples of stars both with and without detected planets
behave quite similarly, although there is an average overabundance of
about 0.15 dex in the total planet-host stars with respect to the
comparison sample ($\langle$[C/H]$\rangle$$_{P}$ = 0.10, $\sigma =
0.16$, ${\rm RMS} = 0.19$ and $\langle$[C/H]$\rangle$$_{C}$ = -0.06,
$\sigma = 0.18$, ${\rm RMS} = 0.19$). Since targets with planets are
on average more metal-rich than the stars of comparison sample, their
abundance distributions correspond to the extensions of the comparison
sample trends at high metallicity (see Fig. \ref{histograma}). Such a trend
supports the primordial scenario as an explanation of
the overmetallicity of planet-host stars. 
C abundances present a bimodality for metallicities
lower than solar, due to the average overabundance of thick disc stars
in comparison with the thin disk stars \citep[see][]{neves}. [C/Fe]
clearly decreases with [Fe/H] in the metallicity range -0.8 $<$ [Fe/H]
$<$ -0.2, but for higher metallicities this ratio is more flatenned.
This flatenning of the [C/Fe] ratios was also found by other authors
\citep[e.g.][]{Sadakane} but in other works a monotonic decrease of
[C/Fe] with metallicity was reported \citep[e.g.][]{andersson,
alex_carb}. Since we do not observe differences between 
the samples with and without detected planets, 
this behaviour must be evidence of
the chemical evolution of the Galactic disk.\\

\subsection{Oxygen}

There are several indicators to measure oxygen abundances: the near-IR
OI triplet at $\lambda$ 7771-5 \AA{}, the forbidden lines of
[\ion{O}{1}] at $\lambda$ 6300 \AA{} and $\lambda$ 6363 \AA{} and the
near-UV OH lines at $\lambda$ 3100 \AA{}. Ecuvillon et al. (2006) made
a comparative study of the three indicators in a sample of stars with
and without detected planets and found good agreement between the
[O/H] ratios from forbidden and OH lines, while the NLTE triplet shows
a systematically lower abundance. Unfortunately, only forbidden line
is available in HARPS spectra, so we used this indicator to obtain
oxygen LTE abundances, since it is well known that this indicator is
not significantly affected by deviations from LTE
\citep[e.g.][]{kiselman}.\\ 

  \begin{figure*}[ht]
   \centering
   \includegraphics[width=7.8cm]{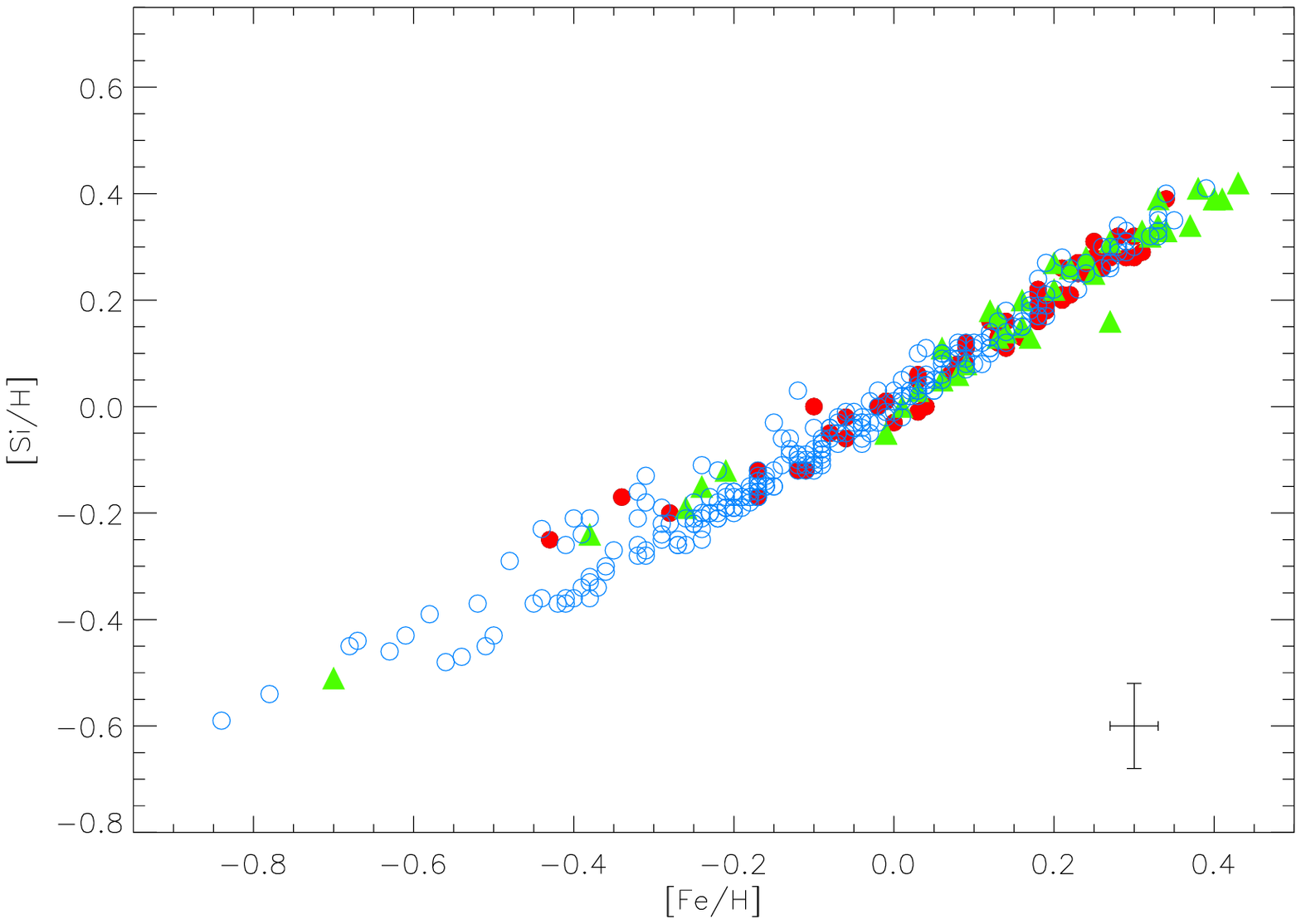}
   \includegraphics[width=7.8cm]{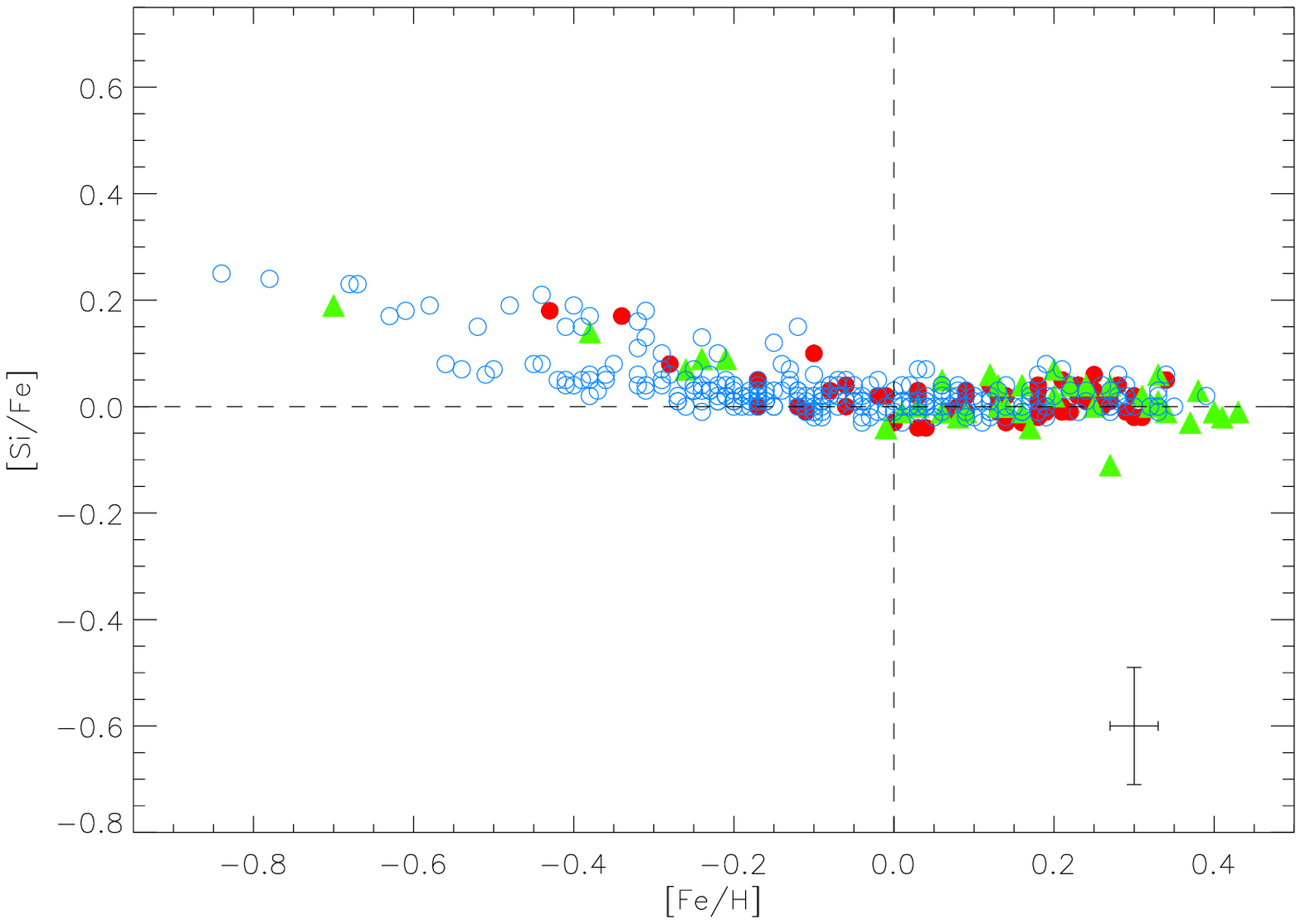}
      \caption{[Si/H] vs [Fe/H] and [Si/Fe] vs [Fe/H] for stars with (red filled circles) and without (blue open circles) detected planets from the HARPS GTO sample. Green triangles are stars with planets from other surveys.}
         \label{sil}
   \end{figure*}

The spectral region around this feature has telluric lines which
can be blended with the [\ion{O}{1}] line in some stars. So we made a
detailed observation of the spectra to remove these objets from the
sample in order to avoid wrong values of the O abundance. This,
together with the limitation on $T_{\rm eff}$, makes a final sample of 69
and 270 stars with and without detected planets from HARPS, and 31
stars with planets from other surveys. This line is also blended with
a \ion{Ni}{1} absorption at $\lambda$ 6300.399 \AA{}
\citep{lambert,allende}, so we estimated the EW of the Ni line 
using the \texttt{ewfind} driver of MOOG \citep{sneden}. Ni abundances for HARPS
stars were taken from \citet{neves}. 
For the additional sample stars we calculate
Ni abundances in the same way as in \citet{neves}. The oxygen
contribution has been obtained by subtracting the Ni EW from
the measured EW of whole 6300.23 \AA{} feature. The
wavelengths, excitation energies of the lower leves and oscillator
strengths of the \ion{Ni}{1} absorption were taken form
\citet{allende}, while the adopted atomic data for [\ion{O}{1}] are
from \citet{lambert}. The $\log gf$ value of the [\ion{O}{1}]
line was slightly modified in order to obtain log
$\epsilon$(O)$_{\sun}$ = 8.74 \citep{Nissen}, which is the solar
value used for the differential analysis (see Table \ref{lineas}). We
also calculated solar O abundance using the solar Harp spectrum
(daytime sky spectrum) and the same model we used in Sect.~\ref{seccarbon},
obtaining log $\epsilon$(O)$_{\sun}$ = 8.60, a quite lower value. In
Fig. \ref{niquel} we can see the effect of Ni in oxygen abundances
which becomes greater for higher Ni abundances, as we might expect.\\

In Fig. \ref{oxig} we show the [O/H] and [O/Fe] ratios as a function
of [Fe/H]. There appear to be no clear differences between stars with and
without detected planets. This result is in disagreement with Si enrichment in stars with planets with respect to stars without known planets found by \citet{Robinson}, since they would also expect to find an O enrichment in these stars. However, there is an average overabundance
of about 0.13 dex in the planet hosts with respect to the
comparison sample ($\langle$[O/H]$\rangle$$_{P}$ = 0.05, $\sigma =
0.16$, ${\rm RMS} = 0.17$ and $\langle$[O/H]$\rangle$$_{C}$ = -0.08,
$\sigma = 0.17$, ${\rm RMS} = 0.19$). As mentioned in 
Sect.~\ref{seccarbon}, the
abundance distributions of stars with planets correspond to the
extensions of the comparison sample trends at high [Fe/H] (see Fig.
\ref{histograma}). [O/Fe] clearly
decreases with [Fe/H] in the metallicity range 
$-0.8<[{\rm Fe}/{\rm H}]<0.0$, 
although this fall is not so steep. This behaviour has been also
reported in previous works \citep{Bensby, alex_oxig}, where [O/Fe] 
showed a monotonic decrease with metallicity, in agreement with 
galactic evolution models. 

\subsection{Magnesium and Silicon\label{secmgsi}}

Mg and Si abundances were calculated using the line list of
\citet{neves}, adding a Mg line at $\lambda$ 6318.72 \AA{}. Solar
values that we used for the differential analysis of the two elements
are log $\epsilon$(Mg)$_{\sun}$ = 7.58 and log $\epsilon$(Si)$_{\sun}$
= 7.55 \citep{anders}. The abundance values obtained from the Harps
1000 spectrum (daytime sky spectrum) are log $\epsilon$(Mg)$_{\sun}$ =
7.54 and log $\epsilon$(Si)$_{\sun}$ = 7.52, slightly lower than the
reference values.\\ 

\begin{figure}[ht!]
   \centering
   \includegraphics[width=6cm]{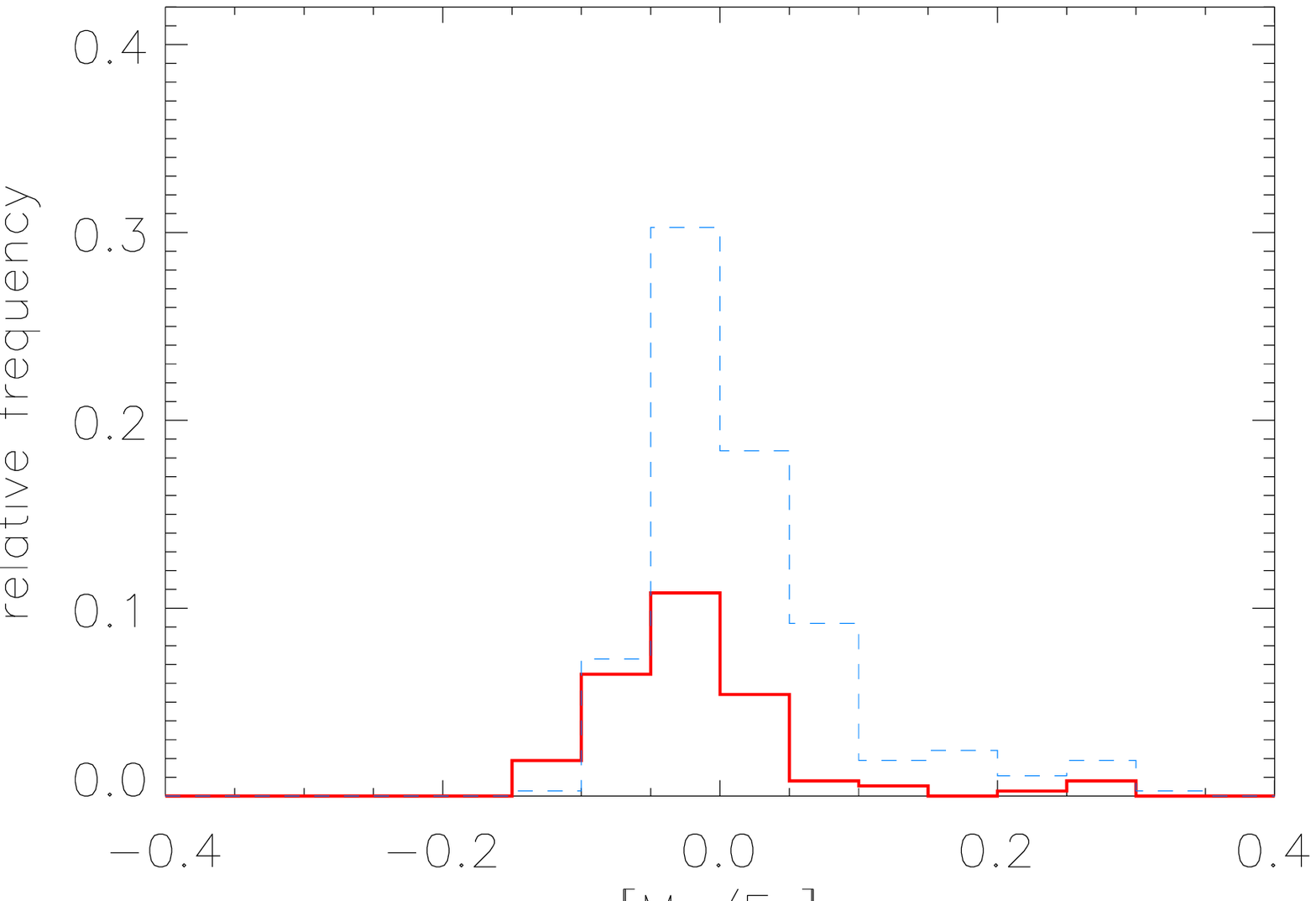}
   \includegraphics[width=6cm]{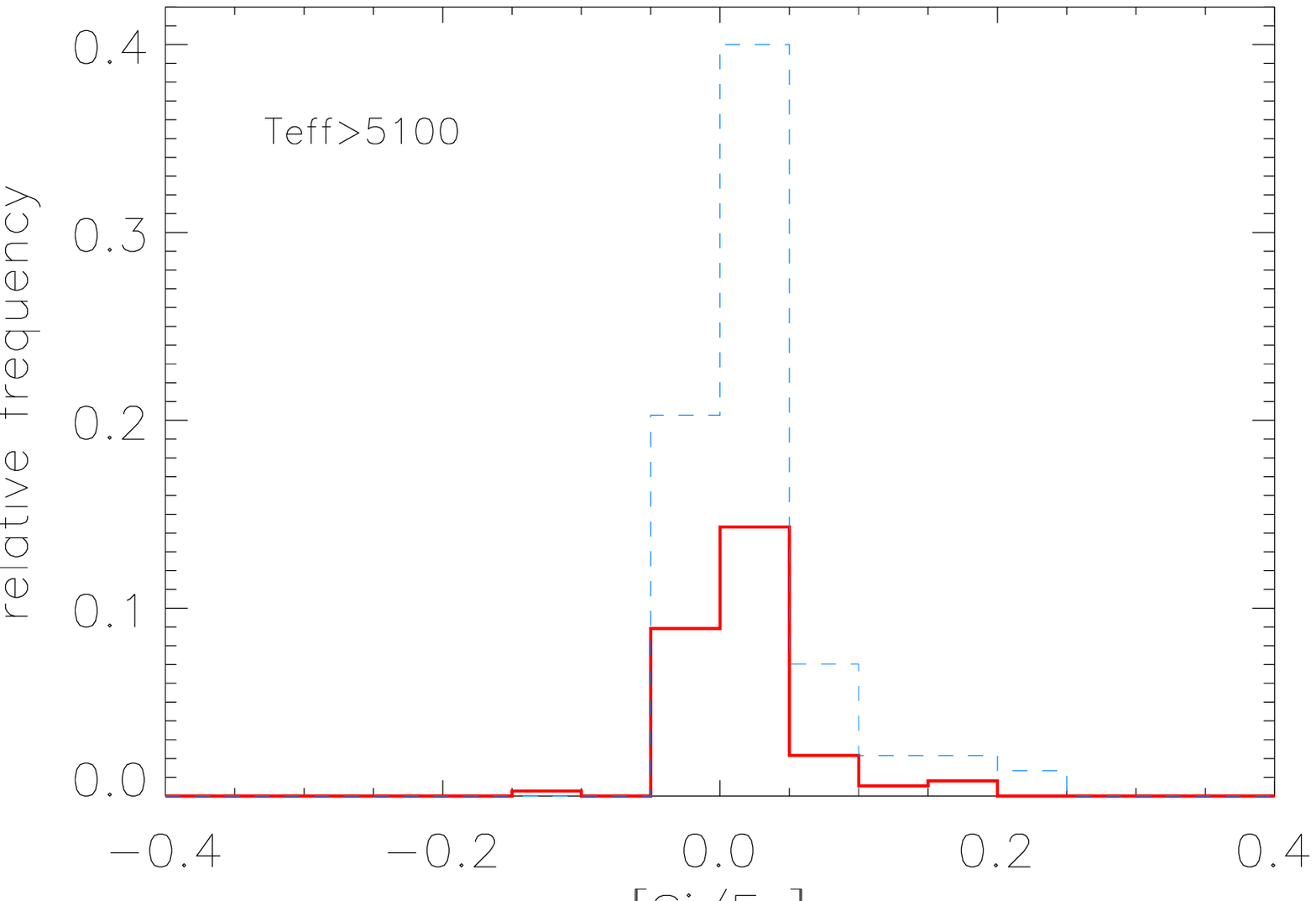}
      \caption{[Si/Fe] and [Mg/Fe] distributions for stars with (red line) and without (blue dashed line) detected planets.}
         \label{histogramaMg}
   \end{figure}

In Fig. \ref{sil} we can see [Si/H] and [Si/Fe] as a function of
[Fe/H]. \citet{Robinson} found clear and significant overabundances of Si in stars with planets with respect to comparison stars. On the other hand \citet{Gonzalez07} reported sistematically lower abundances of this element in the higher metallicity range for stars with planets. However, and in agreement with recent works \citep{neves,jonay}, we do not find significant differences between the stars with
and without detected planets although the average values are 0.19 dex
greater in stars hosting planets ($\langle$[Si/H]$\rangle$$_{P}$ =
0.14, $\sigma=0.17$, ${\rm RMS}= 0.22$ and
$\langle$[Si/H]$\rangle$$_{C}$ = -0.05, $\sigma=0.19$, ${\rm RMS}=
0.20$), again due to the higher metallicity of the planet-host 
sample.
For [Mg/H] there is a similar effect, owing to the same reason
($\langle$[Mg/H]$\rangle$$_{P}$ = 0.10, $\sigma=0.15$, ${\rm RMS}=
0.18$ and $\langle$[Mg/H]$\rangle$$_{C}$ = -0.06, $\sigma = 0.18$,
${\rm RMS} = 0.19$). At subsolar metallicities all stars present high
Mg abundances irrespective of $T_{\rm eff}$. 
However, this is not the
case for 
[Fe/H] $\geq$ 0, where stars without detected planets have higher Mg
abundances, $\langle$[Mg/Fe]$\rangle$$_{P}$ = -0.040, $\sigma=0.04$,
${\rm RMS}=0.06$ and  $\langle$[Mg/Fe]$\rangle$$_{C}$ = -0.014,
$\sigma=0.04$, ${\rm RMS}=0.04$ (see Fig. \ref{magn}), also
for different temperatures. Nevertheless, this effect dissapears when
we take into account only  solar analogs, with 
$5600 < T_{\rm eff} < 5950$~K, perhaps due to the low number 
of stars with planets in this group. 
Therefore, it might be an effect in Mg
abundances due to the presence of planetary companions (see Fig. \ref{histogramaMg}). 
For both elements we observe the same bimodality we found for C abundances at
lower metallicities, owing to the different populations from thin and
thick disk \citep{neves}. [Mg/Fe] and [Si/Fe] ratios show a
decrease for [Fe/H] $<$ 0 but they flatten for higher metallicities as
a consequence of the chemical evolution of the Galaxy.\\ 
 
\section{C/O vs Mg/Si}

\begin{figure}[!h]
   \centering
   \includegraphics[width=8.cm]{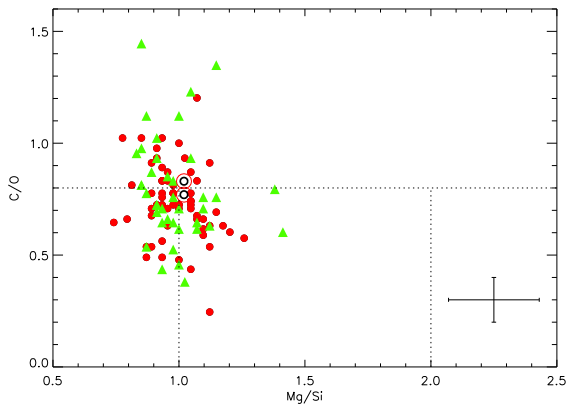}
      \caption{C/O vs Mg/Si for stars with planets from the HARPS GTO sample (red filled circles). Green triangles are stars with planets from other surveys.}
         \label{planetas}
   \end{figure}

In Fig. \ref{planetas}, C/O ratios as a function of Mg/Si are
presented for different temperature ranges. These ratios are
calculated as:
\begin{equation}
{\rm A/B} = N_{\rm A}/N_{\rm B} = 10^{\log \epsilon({\rm A})}/10^{\log
\epsilon({\rm B})}
\label{eqAR}
\end{equation}
where $\log \epsilon ({\rm A})$ and  $\log
\epsilon ({\rm B})$ are the
absolute abundances, so they are not dependent on solar reference
abundances. In our sample, 34\% of stars with known planets have C/O 
values greater than 0.8, which means that Si will exist primarily as SiC 
(see Sect. \ref{planet-comp}). On the other hand, 66\% 
of stars with known planets have C/O values lower than 0.8 and Si will be present in rock-forming minerals as the SiO$_{2}$ structural unit. In these cases, silicate mineralogy will be controlled by Mg/Si ratio. 52\% of these stars (with C/O $<$ 0.8) present Mg/Si ratios between 1 and 2, similar to the solar ratio, while 48\% have ratios lower than 1. We do not find any star with Mg/Si $>$ 2. If we take into account all stars, irrespective of their C/O value, these percentages are similar (see Table \ref{tabla_CO}).\\

\begin{figure}[!h]
   \centering
   \includegraphics[width=8.cm]{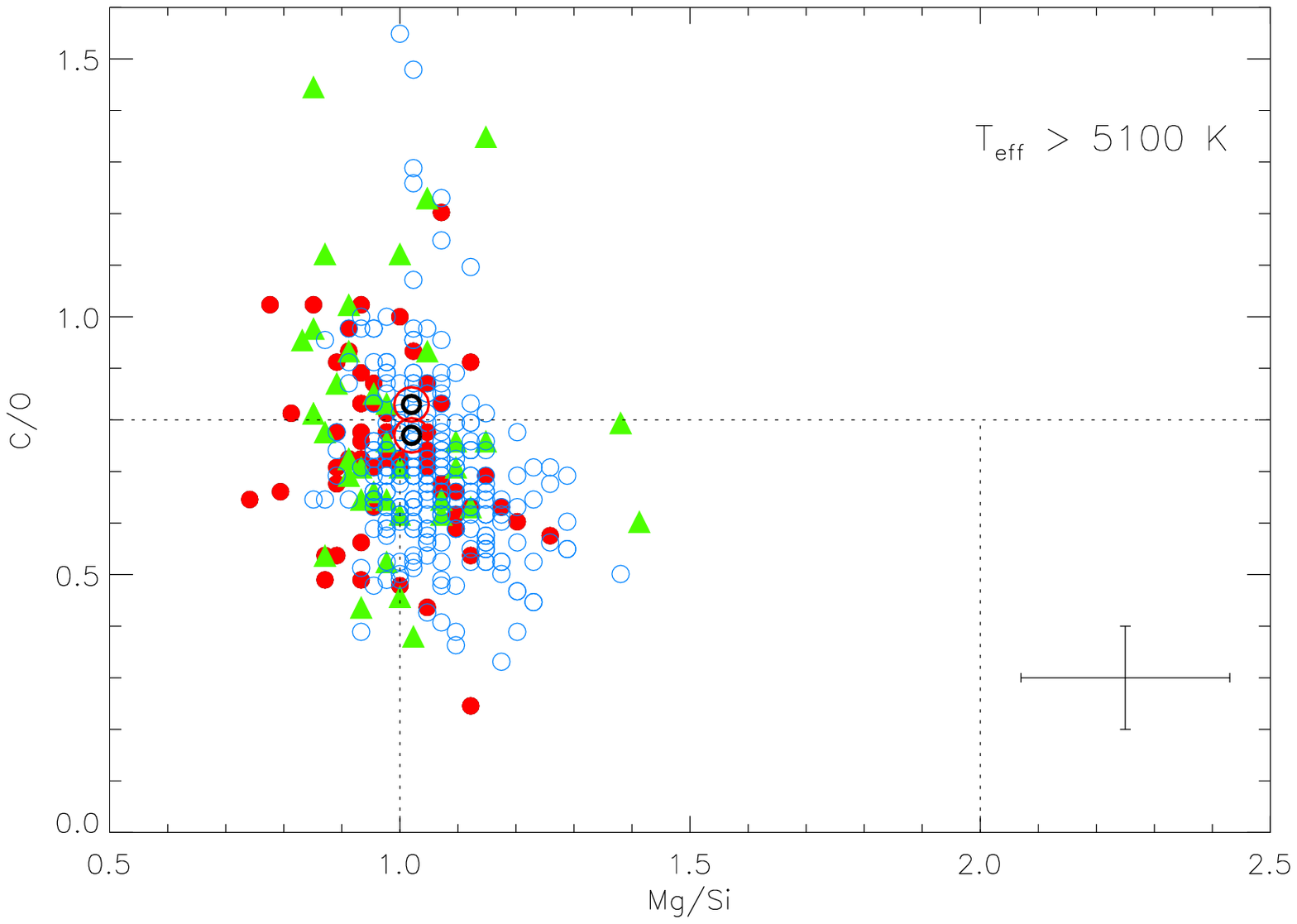}
   \includegraphics[width=8.cm]{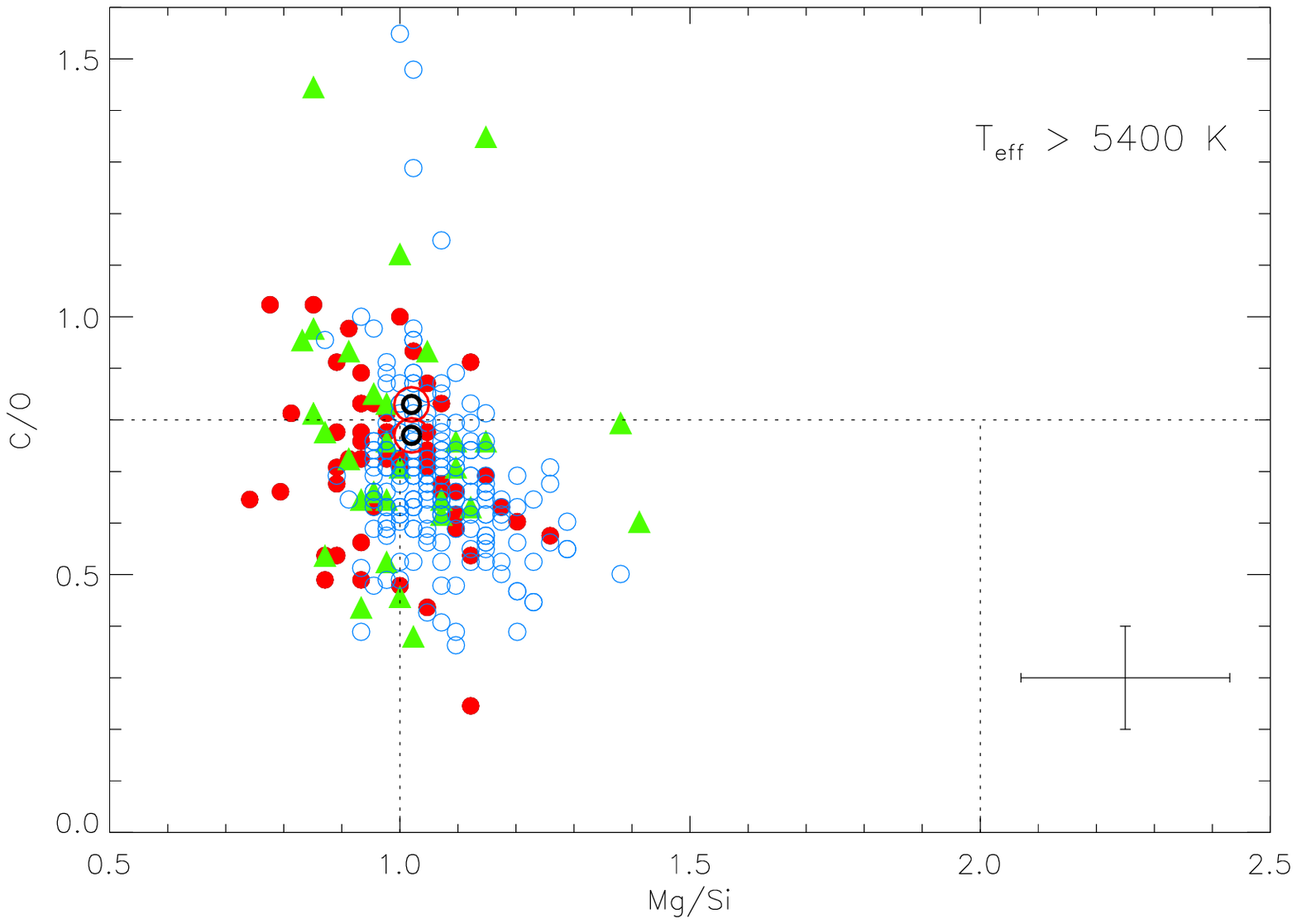}
   \includegraphics[width=8.cm]{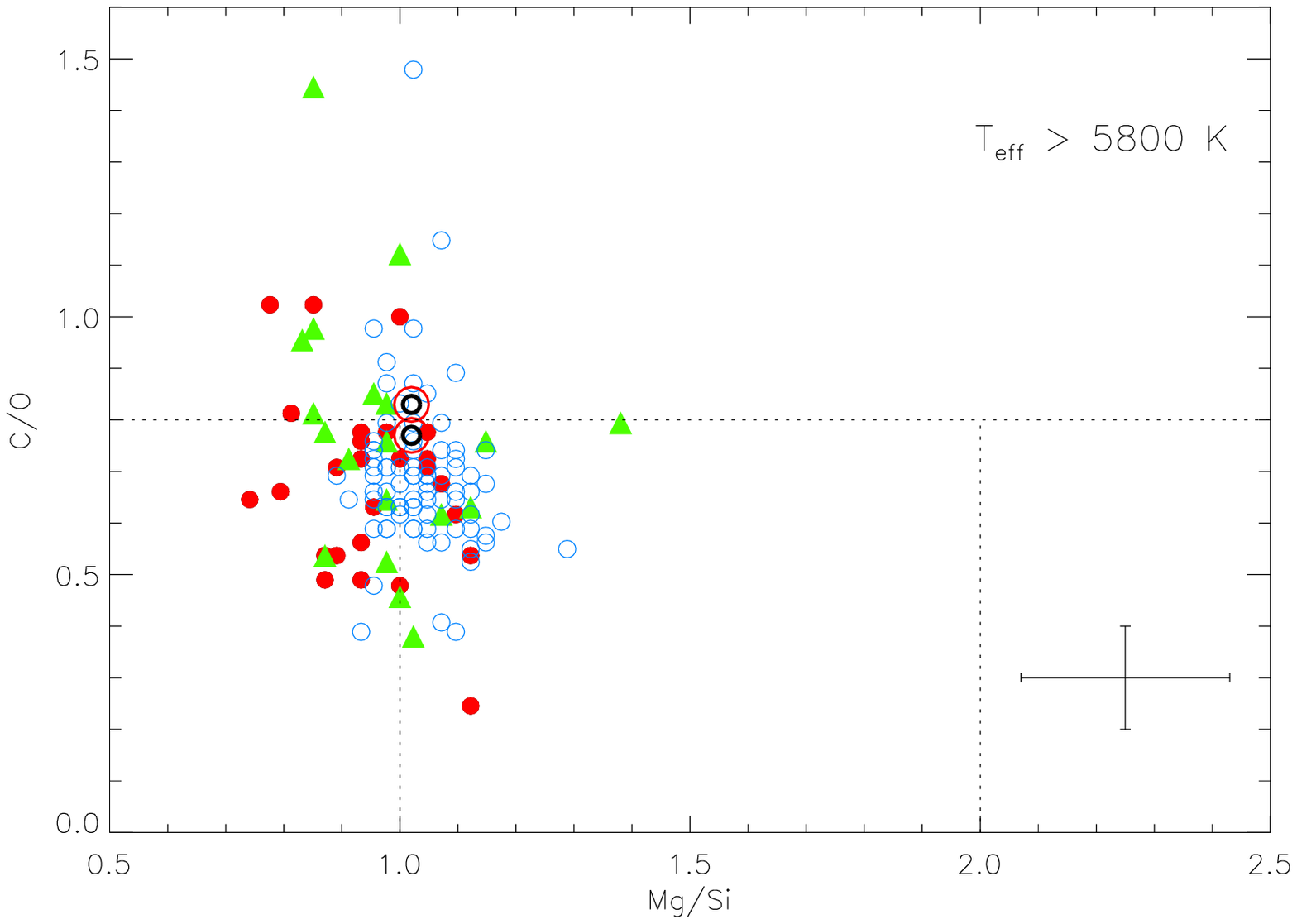}
      \caption{C/O vs Mg/Si for stars with (red filled circles) and without (blue open circles) detected planets from the HARPS GTO sample. Green triangles are stars with planets from other surveys.}
         \label{CO}
   \end{figure}

Comparison sample stars are shifted towards higher Mg/Si
ratios (see Fig. \ref{CO}), since they present higher Mg abundances as 
mentioned in Sect.~\ref{secmgsi}. We do not find any significant effect
related to the effective temperature of the stars (see Fig. \ref{CO}). Both Atlas
and Harps solar ratios are represented in the plots. Mg/Si ratios are
equal for both spectra although C/O ratio is a little greater for Harps
spectrum. In any case, this value is in the lowest limit of C-rich
systems.\\

\begin{center}
\begin{table*}[ht]
\caption{C/O and Mg/Si distributions for stars with planets}
\label{tabla_CO}
\centering
\begin{tabular}{ccl}
\hline
\noalign{\medskip} 
Ratio &  Percentage & Principal Composition\\
\noalign{\medskip} 
\hline
\hline
\noalign{\medskip} 
C/O $>$ 0.8 & 34\% & graphite, TiC and solid Si as SiC \\
C/O $<$ 0.8 & 66\% & solid Si as SiO$_{4}$$^{4-}$ or SiO$_{2}$ \\
\noalign{\medskip} 
\hline
\noalign{\medskip} 
Mg/Si $<$ 1 & 56\% & pyroxene, metallic Fe and excess Si as feldspars\\
1 $<$ Mg/Si $<$ 2 & 44\% & equal pyroxene and olivine \\
Mg/Si $>$2 & 0\% & olivine and excess Mg as MgO \\
\hline
\end{tabular}
\end{table*}
\end{center}

The errors in the abundance ratios C/O and Mg/Si were estimated by evaluating 
an increase or a decrease in the $\log \epsilon ({\rm A})-\log \epsilon ({\rm B})$ abundance
ratio, due to the relative error, using the Eq.~\ref{eqAR}.\\

\subsection{Terrestrial Planet Compositions\label{planet-comp}}

The wide variety of host star compositions determined in this study
will presumably result in a diverse range of compositions of solid
material available for terrestrial planet formation. As previously
discussed by \cite{bond_sim}, under the assumption of equilibrium
those systems with a C/O value above 0.8 will contain carbide-rich
phases (such as graphite, SiC and TiC) in the innermost regions of the
disk. Metallic Fe and Mg-silicates such as olivine (Mg$_{2}$SiO$_{4}$)
and pyroxene (MgSiO$_{3}$) are also present and are located further
from the host star. Terrestrial planets forming in these planetary
systems are expected to be C-rich, containing significant amounts of C
in addition to Si, Fe, Mg and O. 

For systems with a C/O value below 0.8, Si will be present in the
solid form primarily as SiO$_{4}$$^{4-}$ or SiO$_{2}$, predominantly
forming Mg-silicates. The exact composition of the Mg-silicates is
controlled by the Mg/Si value. For systems with a Mg/Si value between
1 and 2, the silicates present are predominately olivine and pyroxene
in a condensation sequence closely resembling Solar. This is expected
to result in the production of terrestrial planets similar in
composition to that of Earth (in that their composition will be
dominated by O, Fe, Mg and Si, with small amounts of Ca and Al also
present). 

However, 56\% of all planetary host stars in this study have a
Mg/Si value less than 1. For such a composition, the solid component
of the disk is dominated by approximately equal amounts of pyroxene
and metallic Fe with minimal amounts of olivine present. Feldspars are
also likely to be present as all available Mg is partioned into
pyroxene, leaving excess Si available to form other silicate species.
This is expected to result in the production of terrestrial planets
that can be best described as being Si-rich Earths. They will still be
dominated by O, Fe, Mg and Si and contain minor amounts of other
elements such as Ca and Al. However, their bulk Si content is expected
to be well above any value previously observed for a planetary body.
Note that for this study, the high-Si planetary compositions are due
to the fact that there is an excess of Si compared to Mg within the
disk system and does not necessarily imply an elevated Si abundance.
Such an elevated Si content is predicted to produce a quartz-feldspar
rich terrestrial planet with a composition more like that of Earth's
continental crust material than that of Earth's olivine-dominated
mantle. A composition such as this can have drastic implications for
planetary processes such as plate tectonics and atmospheric
composition. For example, volcanism on a Si-rich planet is expected to
be intermediate to felsic in composition (i.e. $>$52\% silica by
weight) due to the potentially high SiO$_{2}$ content of the planet
itself, producing igneous species such as andesite, rhyolite and
granite. Eruptions may also be more explosive in nature due to the
high viscosity of SiO$_{2}$-rich magma trapping volatiles within the
magma. On Earth, such eruptions are commonly observed at convergent
tectonic plate margins (i.e. subduction zones) (for intermediate
compositions) and above intra-plate hot spots (for felsic
compositions). Mount Pinatubo is a well-known example of an
intermediate composition volcano while the Long Valley Caldera in CA,
USA, is an example of a felsic eruption. Although the full
implications of the compositional variations described here still
require detailed study, it is clear that a diverse range  of
terrestrial planets are likely to exist in extrasolar planetary
systems. 

\subsection{Planet Formation\label{planet-form}}

\begin{figure}[ht]
   \centering
   \includegraphics[width=8.cm]{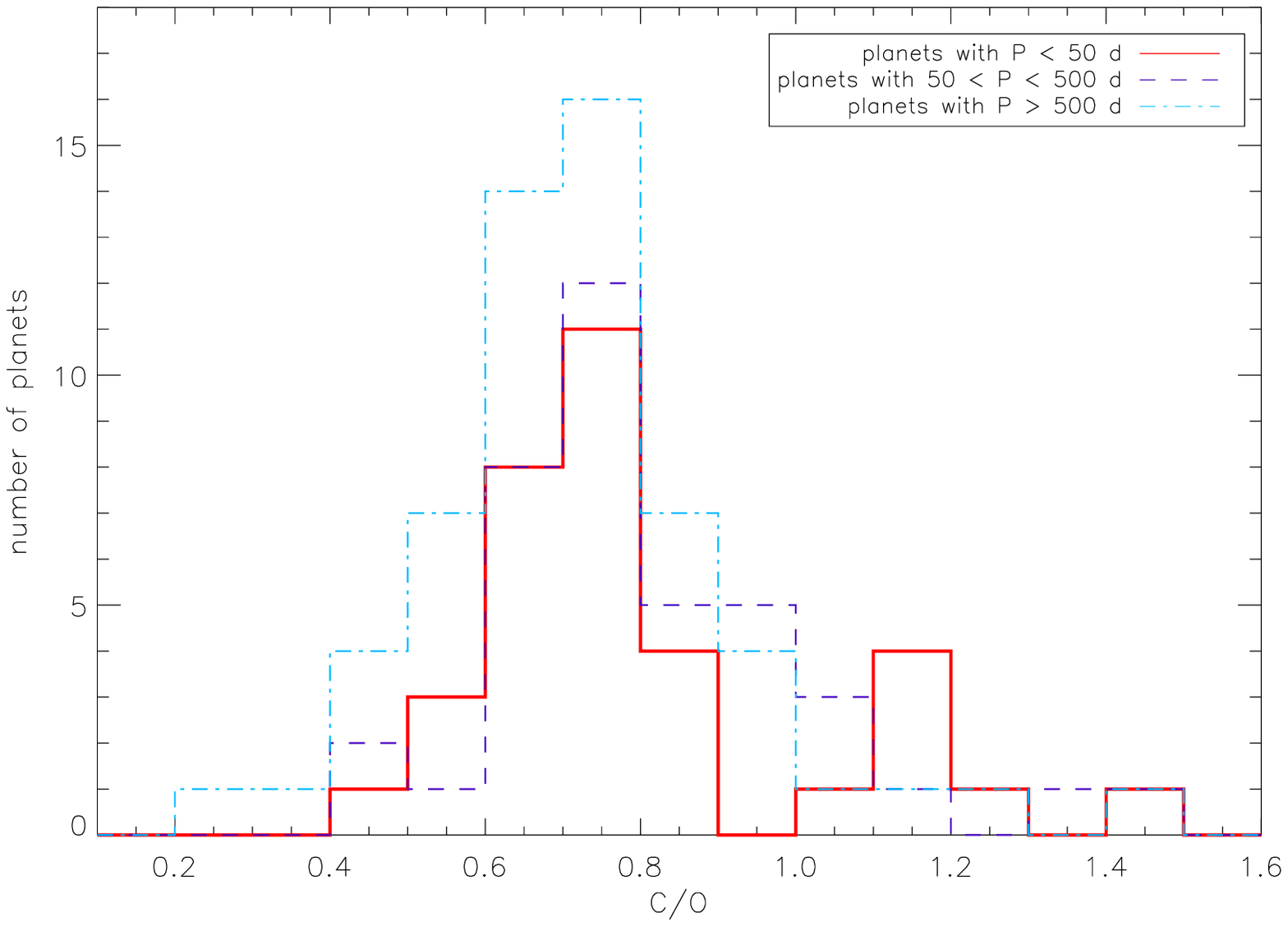}
   \includegraphics[width=8.cm]{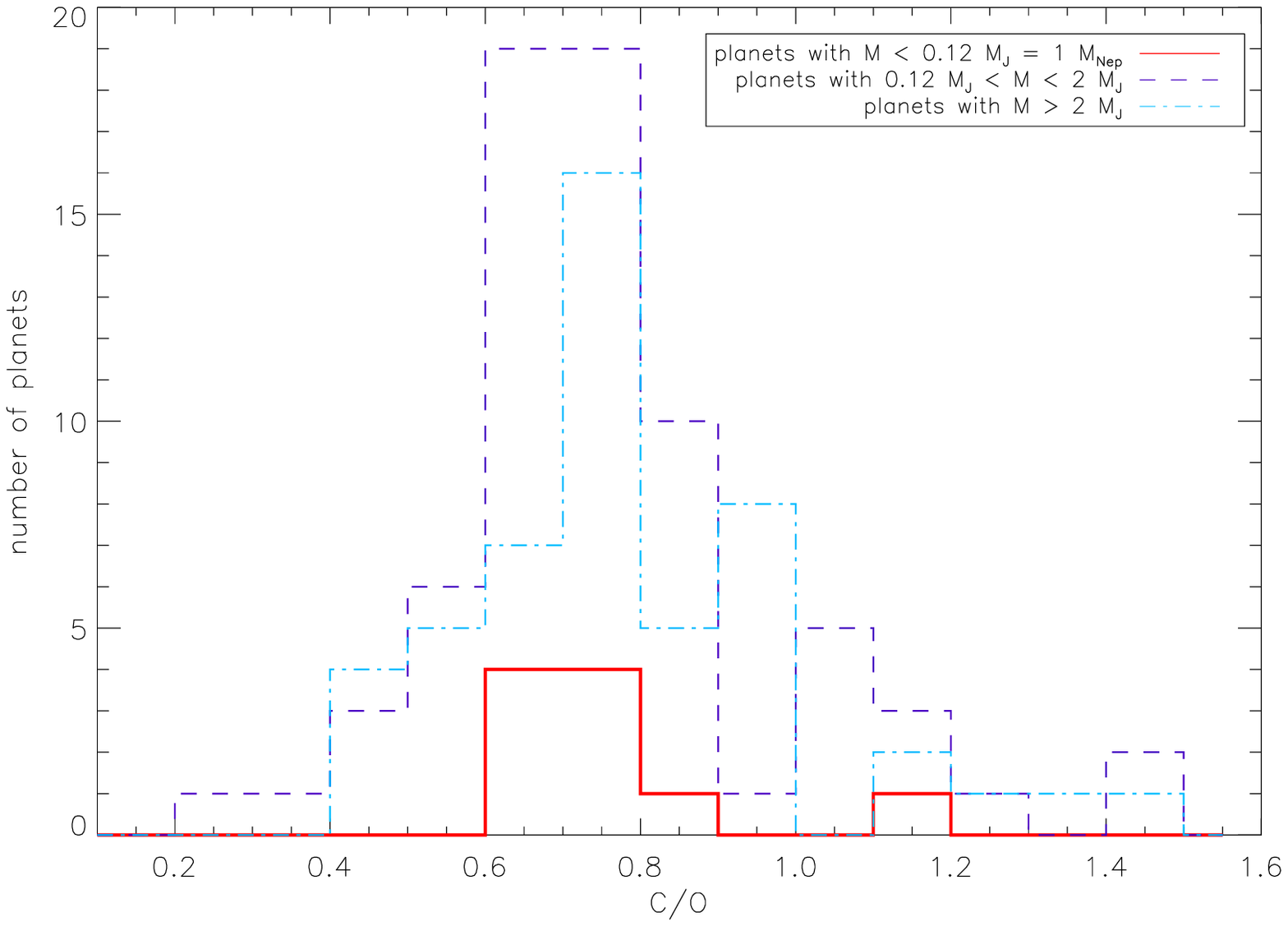}
      \caption{C/O distributions of planet host stars for different period and mass ranges of their companions.}
         \label{CO_hist}
   \end{figure}

It has been previously suggested \citep[e.g.][]{bond_sim} that
planetary systems with C/O values above 0.8 may possess an alternative
mass distribution profile for solid material, potentially making it
easier either for giant planets to form closer to the host star than
previously expected or for terrestrial planets to form in the inner
regions of the disk. However, we find no evidence of any trends with
C/O values for either planetary period, semi-major axis or mass (see
Fig. \ref{CO_hist}). As such, it appears that any effects of an
alternative solid mass distribution due to high concentrations of
refractory C-rich material are not preserved in the architecture of
the system. This is believed to be due to the fact that \cite{bond_sim} only considered equilibrium-driven condensation and did not include the effects of disequilibrium or the migration and radial mixing of material within the disk. Simulations addressing this issue are in progress.
It should be noted, however, that we are still only able
to detect giant planets. This conclusion may be not hold for
extrasolar terrestrial planets which require significantly smaller amounts of solid material.\\

\section{Conclusions}

We present a detailed study of C, O, Mg and Si abundances for a sample
of 100 and 270 stars with and without known giant planets with
effective temperatures between 5100 K and 6500 K, with the aim of studying
the mineralogical composition of terrestrial planets that could have
formed in those systems.\\ 

We do not observe any special difference between abundances of stars
with and without detected planets for C, O and Si. However, we find
higher Mg abundances for stars without detected planets making the
Mg/Si ratio greater in those stars. This effect is not so clear for
solar analogs but the number of stars is not large enough to discard a
possible effect due to the presence of planets.\\ 

C/O and Mg/Si ratios were obtained to study the mineralogy the
possible planets that could have formed around these stars. 34\% of
stars with known planets have C/O values greater than 0.8, so there is
a big fraction of C-rich systems, very different from our Solar
System. On the other hand, 56\% of stars with known planets present
Mg/Si values lower than 1, so these systems are more probably to host
Si-rich earths, with a Si excess much greater than any value
previously observed for a planetary body. This can have extreme
implications for processes as plate tectonics or volcanism. We also
found stars very similar to our Sun but it is clear that a wide
variety of planets will probably exist within extrasolar planetary
systems.\\ 

\acknowledgments

E.D.M, J.I.G.H. and G.I. would like to thank financial
support from the Spanish Ministry project MICINN AYA2008-04874.
J.I.G.H. acknowledges financial support from the Spanish Ministry 
project MICINN AYA2008-00695 and also from the Spanish Ministry of
Science and Innovation (MICINN) under the 2009 Juan de la Cierva
Programme.\\ 
N.C.S. would like to thank the support by the 
European Research Council/European Community under the FP7 through a 
Starting Grant, as well from Funda\c{c}\~ao para a Ci\^encia e a
Tecnologia (FCT), Portugal, through a Ci\^encia\,2007 
contract funded by FCT/MCTES (Portugal) and POPH/FSE (EC), 
and in the form of grant reference PTDC/CTE-AST/098528/2008
from FCT/MCTES.\\
This work has also made use of
the IRAF facility, and the Encyclopaedia of extrasolar planets.






\appendix

\section{Appendix material}

\begin{table}
\caption{Stars with planets from the HARPS GTO survey.}
\label{lista_planetas1}
\begin{tabular}{lcccrrrrrrrr}
\noalign{\medskip} 
\hline
\noalign{\medskip} 
Star & T$_{\rm eff}$ & log \textit{g} & $\xi_{t}$ & [Fe/H] & [O/H] & [Ni/H] & [C/H]& [Mg/H] & [Si/H]& C/O & Mg/Si\\
& K & & km s$^{-1}$ & & & & & & & &\\
\noalign{\medskip} 
\hline
\hline
\noalign{\medskip} 
    HD142  &   6403.  & 4.62  &  1.74  &  0.09  & -0.02  &  0.01  &  0.17  & -0.03  &  0.11  &  1.02  &  0.78 \\
   HD1237  &   5514.  & 4.50  &  1.09  &  0.07  & -0.18  &  0.06  & -0.01  & -0.01  &  0.06  &  0.98  &  0.91 \\
   HD2638  &   5198.  & 4.43  &  0.74  &  0.12  & -0.01  &  0.15  &  0.18  &  0.10  &  0.16  &  1.02  &  0.93 \\
   HD4308  &   5644.  & 4.38  &  0.90  & -0.34  & -0.10  & -0.30  & -0.14  & -0.12  & -0.17  &  0.60  &  1.20 \\
  HD10647  &   6218.  & 4.62  &  1.22  &  0.00  &  0.06  & -0.05  & -0.07  & -0.09  & -0.03  &  0.49  &  0.93 \\
 HD11964A  &   5332.  & 3.90  &  0.99  &  0.08  &  0.06  &  0.07  &  0.04  &  0.10  &  0.08  &  0.63  &  1.12 \\
  HD16141  &   5806.  & 4.19  &  1.11  &  0.16  &  0.08  &  0.15  &  0.09  &  0.13  &  0.13  &  0.68  &  1.07 \\
  HD16417  &   5841.  & 4.16  &  1.18  &  0.13  &  0.09  &  0.15  &  0.06  &  0.14  &  0.13  &  0.62  &  1.10 \\
  HD17051  &   6227.  & 4.53  &  1.29  &  0.19  &  0.07  &  0.18  &  0.07  &  0.06  &  0.19  &  0.66  &  0.79 \\
  HD19994  &   6289.  & 4.48  &  1.72  &  0.24  &  0.30  &  0.27  &  0.29  &  0.09  &  0.25  &  0.65  &  0.74 \\
  HD20782  &   5774.  & 4.37  &  1.00  & -0.06  & -0.12  & -0.09  & -0.07  & -0.07  & -0.06  &  0.74  &  1.05 \\
  HD22049  &   5153.  & 4.53  &  0.90  & -0.11  & -0.12  & -0.15  &  0.00  & -0.17  & -0.12  &  0.87  &  0.95 \\
  HD23079  &   5980.  & 4.48  &  1.12  & -0.12  & -0.18  & -0.15  & -0.14  & -0.13  & -0.12  &  0.72  &  1.05 \\
  HD28185  &   5667.  & 4.42  &  0.94  &  0.21  &  0.08  &  0.28  &  0.22  &  0.18  &  0.26  &  0.91  &  0.89 \\
  HD39091  &   6003.  & 4.42  &  1.12  &  0.09  &  0.01  &  0.10  &  0.04  &  0.07  &  0.08  &  0.71  &  1.05 \\
  HD45364  &   5434.  & 4.38  &  0.71  & -0.17  & -0.17  & -0.16  & -0.17  & -0.12  & -0.12  &  0.66  &  1.07 \\
  HD47186  &   5675.  & 4.36  &  0.93  &  0.23  &  0.09  &  0.30  &  0.19  &  0.22  &  0.27  &  0.83  &  0.95 \\
  HD52265  &   6136.  & 4.36  &  1.32  &  0.21  &  0.06  &  0.22  &  0.13  &  0.15  &  0.21  &  0.78  &  0.93 \\
  HD65216  &   5612.  & 4.44  &  0.78  & -0.17  & -0.22  & -0.21  & -0.20  & -0.14  & -0.17  &  0.69  &  1.15 \\
  HD66428  &   5705.  & 4.31  &  0.96  &  0.25  &  0.09  &  0.32  &  0.24  &  0.26  &  0.28  &  0.93  &  1.02 \\
  HD69830  &   5402.  & 4.40  &  0.80  & -0.06  & -0.09  & -0.05  & -0.06  & -0.05  & -0.02  &  0.71  &  1.00 \\
  HD70642  &   5668.  & 4.40  &  0.82  &  0.18  &  0.10  &  0.23  &  0.11  &  0.14  &  0.22  &  0.68  &  0.89 \\
  HD73256  &   5526.  & 4.42  &  1.11  &  0.23  &  0.10  &  0.27  &  0.17  &  0.19  &  0.27  &  0.78  &  0.89 \\
  HD75289  &   6161.  & 4.37  &  1.29  &  0.30  &  0.08  &  0.30  &  0.12  &  0.22  &  0.28  &  0.72  &  0.93 \\
  HD82943  &   5989.  & 4.43  &  1.10  &  0.26  &  0.16  &  0.30  &  0.19  &  0.18  &  0.26  &  0.71  &  0.89 \\
  HD83443  &   5511.  & 4.43  &  0.93  &  0.34  &  0.24  &  0.44  &  0.31  &  0.35  &  0.39  &  0.78  &  0.98 \\
  HD92788  &   5744.  & 4.39  &  0.95  &  0.27  &  0.09  &  0.33  &  0.22  &  0.22  &  0.28  &  0.89  &  0.93 \\
  HD93083  &   5105.  & 4.43  &  0.94  &  0.09  &  0.12  &  0.10  &  0.27  &  0.05  &  0.12  &  0.93  &  0.91 \\
 HD100777  &   5536.  & 4.33  &  0.81  &  0.25  &  0.15  &  0.32  &  0.24  &  0.23  &  0.27  &  0.81  &  0.98 \\
 HD101930  &   5164.  & 4.40  &  0.91  &  0.13  &  0.14  &  0.15  &  0.19  &  0.14  &  0.15  &  0.74  &  1.05 \\
 HD102117  &   5657.  & 4.31  &  0.99  &  0.28  &  0.13  &  0.33  &  0.23  &  0.26  &  0.32  &  0.83  &  0.93 \\
 HD107148  &   5805.  & 4.40  &  0.93  &  0.31  &  0.15  &  0.38  &  0.22  &  0.28  &  0.29  &  0.78  &  1.05 \\
 HD108147  &   6260.  & 4.47  &  1.30  &  0.18  &  0.15  &  0.15  &  0.06  &  0.07  &  0.16  &  0.54  &  0.87 \\
 HD111232  &   5460.  & 4.43  &  0.62  & -0.43  & -0.11  & -0.39  & -0.17  & -0.18  & -0.25  &  0.58  &  1.26 \\
 HD114729  &   5844.  & 4.19  &  1.23  & -0.28  & -0.11  & -0.29  & -0.20  & -0.18  & -0.20  &  0.54  &  1.12 \\
 HD114783  &   5133.  & 4.42  &  0.88  &  0.03  & -0.08  &  0.07  &  0.18  &  0.06  &  0.06  &  1.20  &  1.07 \\
 HD117207  &   5667.  & 4.32  &  1.01  &  0.22  &  0.13  &  0.24  &  0.17  &  0.18  &  0.21  &  0.72  &  1.00 \\
 HD117618  &   5990.  & 4.41  &  1.13  &  0.03  & -0.07  &  0.04  &  0.00  &  0.01  &  0.05  &  0.78  &  0.98 \\
 HD121504  &   6022.  & 4.49  &  1.12  &  0.14  &  0.02  &  0.12  &  0.00  &  0.06  &  0.11  &  0.63  &  0.95 \\
 HD130322  &   5365.  & 4.37  &  0.90  & -0.02  & -0.08  & -0.02  & -0.05  & -0.05  &  0.00  &  0.71  &  0.95 \\
 HD134987  &   5740.  & 4.30  &  1.08  &  0.25  &  0.18  &  0.32  &  0.28  &  0.25  &  0.31  &  0.83  &  0.93 \\
 HD141937  &   5893.  & 4.45  &  1.00  &  0.13  &  0.00  &  0.12  &  0.06  &  0.06  &  0.12  &  0.76  &  0.93 \\
HD142022A  &   5508.  & 4.35  &  0.83  &  0.19  & -0.01  &  0.20  &  0.13  &  0.20  &  0.18  &  0.91  &  1.12 \\
\hline
\end{tabular}
\end{table}

\begin{table}
\caption{Stars with planets from the HARPS GTO survey.}
\label{lista_planetas2}
\begin{tabular}{lcccrrrrrrrr}
\noalign{\medskip} 
\hline
\noalign{\medskip} 
Star & T$_{\rm eff}$ & log \textit{g} & $\xi_{t}$ & [Fe/H] & [O/H] & [Ni/H] & [C/H]& [Mg/H] & [Si/H]& C/O & Mg/Si\\
& K & & km s$^{-1}$ & & & & & & & &\\
\noalign{\medskip} 
\hline
\hline
\noalign{\medskip} 
 HD147513  &   5858.  & 4.50  &  1.03  &  0.03  &  0.25  &  0.09  & -0.18  &  0.01  & -0.01  &  0.25  &  1.12 \\
 HD159868  &   5558.  & 3.96  &  1.02  & -0.08  & -0.08  & -0.09  & -0.13  & -0.04  & -0.05  &  0.59  &  1.10 \\
 HD160691  &   5780.  & 4.27  &  1.09  &  0.30  &  0.22  &  0.35  &  0.26  &  0.28  &  0.32  &  0.72  &  0.98 \\
 HD168746  &   5568.  & 4.33  &  0.81  & -0.10  &  0.01  & -0.08  & -0.01  &  0.04  &  0.00  &  0.63  &  1.17 \\
 HD169830  &   6361.  & 4.21  &  1.56  &  0.18  &  0.19  &  0.18  &  0.12  &  0.13  &  0.19  &  0.56  &  0.93 \\
 HD179949  &   6287.  & 4.54  &  1.36  &  0.21  &  0.05  &  0.21  &  0.14  &  0.08  &  0.20  &  0.81  &  0.81 \\
 HD190647  &   5639.  & 4.18  &  0.99  &  0.23  &  0.43  &  0.27  &  0.25  &  0.24  &  0.25  &  0.44  &  1.05 \\
 HD196050  &   5917.  & 4.32  &  1.21  &  0.23  &  0.36  &  0.29  &  0.22  &  0.23  &  0.26  &  0.48  &  1.00 \\
 HD202206  &   5757.  & 4.47  &  1.01  &  0.29  &  0.09  &  0.33  &  0.13  &  0.21  &  0.28  &  0.72  &  0.91 \\
 HD204313  &   5776.  & 4.38  &  1.00  &  0.18  &  0.11  &  0.22  &  0.18  &  0.15  &  0.19  &  0.78  &  0.98 \\
 HD208487  &   6146.  & 4.48  &  1.24  &  0.08  & -0.14  &  0.06  &  0.04  &  0.04  &  0.07  &  1.00  &  1.00 \\
 HD210277  &   5505.  & 4.30  &  0.86  &  0.18  &  0.14  &  0.21  &  0.24  &  0.21  &  0.21  &  0.83  &  1.07 \\
 HD212301  &   6271.  & 4.55  &  1.29  &  0.18  &  0.18  &  0.19  &  0.09  &  0.09  &  0.17  &  0.54  &  0.89 \\
 HD213240  &   5982.  & 4.27  &  1.25  &  0.14  &  0.07  &  0.16  &  0.11  &  0.13  &  0.16  &  0.72  &  1.00 \\
 HD216435  &   6008.  & 4.20  &  1.34  &  0.24  & -0.02  &  0.28  &  0.17  &  0.16  &  0.26  &  1.02  &  0.85 \\
 HD216770  &   5424.  & 4.38  &  0.91  &  0.24  &  0.13  &  0.32  &  0.25  &  0.26  &  0.27  &  0.87  &  1.05 \\
 HD221287  &   6374.  & 4.62  &  1.29  &  0.04  &  0.06  & -0.02  & -0.07  & -0.09  &  0.00  &  0.49  &  0.87 \\
 HD222582  &   5779.  & 4.37  &  1.00  & -0.01  & -0.01  &  0.00  & -0.01  &  0.02  &  0.01  &  0.66  &  1.10 \\
\hline
\end{tabular}
\end{table}

\begin{table}
\caption{Stars with planets from other surveys.}
\label{lista_alex}
\begin{tabular}{lcccrrrrrrrr}
\noalign{\medskip} 
\hline
\noalign{\medskip} 
Star & T$_{\rm eff}$ & log \textit{g} & $\xi_{t}$ & [Fe/H] & [O/H] & [Ni/H] & [C/H]& [Mg/H] & [Si/H]& C/O & Mg/Si\\
& K & & km s$^{-1}$ & & & & & & & &\\
\noalign{\medskip} 
\hline
\hline
\noalign{\medskip} 

   HD2039  &   5976 &  4.45  &  1.26  &  0.32  &  0.11  &  0.35  &  0.20  &  0.22  &  0.32  &  0.81  &  0.85 \\
   HD3651  &   5173 &  4.37  &  0.74  &  0.12  &  0.06  &  0.15  &  0.25  &  0.11  &  0.18  &  1.02  &  0.91 \\
   HD4203  &   5636 &  4.23  &  1.12  &  0.40  &  0.25  &  0.42  &  0.40  &  0.32  &  0.39  &  0.93  &  0.91 \\
   HD8574  &   6151 &  4.51  &  1.45  &  0.06  &  0.16  &  0.04  &  0.00  &  0.02  &  0.05  &  0.46  &  1.00 \\
   HD9826  &   6212 &  4.26  &  1.69  &  0.13  & -0.19  &  0.09  &  0.15  &  0.07  &  0.17  &  1.45  &  0.85 \\
  HD10697  &   5641 &  4.05  &  1.13  &  0.14  & -0.05  &  0.13  &  0.10  &  0.13  &  0.14  &  0.93  &  1.05 \\
  HD13445  &   5163 &  4.52  &  0.72  & -0.24  & -0.34  & -0.26  & -0.07  & -0.16  & -0.15  &  1.23  &  1.05 \\
  HD20367  &   6138 &  4.53  &  1.22  &  0.17  &  0.21  &  0.07  & -0.03  &  0.11  &  0.13  &  0.38  &  1.02 \\
  HD23596  &   6108 &  4.25  &  1.30  &  0.31  &  0.11  &  0.35  &  0.27  &  0.22  &  0.33  &  0.95  &  0.83 \\
  HD30177  &   5588 &  4.29  &  1.08  &  0.38  &  0.24  &  0.42  &  0.33  &  0.31  &  0.41  &  0.81  &  0.85 \\
  HD37124  &   5546 &  4.50  &  0.80  & -0.38  & -0.15  & -0.39  & -0.19  & -0.12  & -0.24  &  0.60  &  1.41 \\
  HD38529  &   5674 &  3.94  &  1.38  &  0.40  &  0.45  &  0.42  &  0.27  &  0.33  &  0.39  &  0.44  &  0.93 \\
  HD41004  &   5242 &  4.35  &  1.01  &  0.16  &  0.07  &  0.16  &  0.19  &  0.12  &  0.20  &  0.87  &  0.89 \\
  HD46375  &   5268 &  4.41  &  0.97  &  0.20  &  0.28  &  0.26  &  0.30  &  0.20  &  0.27  &  0.69  &  0.91 \\
  HD50554  &   6026 &  4.41  &  1.11  &  0.01  & -0.14  & -0.04  & -0.08  & -0.04  &  0.00  &  0.76  &  0.98 \\
  HD72659  &   5995 &  4.30  &  1.42  &  0.03  & -0.01  &  0.01  & -0.02  & -0.01  &  0.03  &  0.65  &  0.98 \\
  HD73526  &   5699 &  4.27  &  1.26  &  0.27  &  0.24  &  0.27  &  0.23  &  0.25  &  0.31  &  0.65  &  0.93 \\
  HD74156  &   6112 &  4.34  &  1.38  &  0.16  &  0.01  &  0.15  &  0.08  &  0.06  &  0.15  &  0.78  &  0.87 \\
  HD75732  &   5279 &  4.37  &  0.98  &  0.33  &  0.07  &  0.39  &  0.30  &  0.30  &  0.39  &  1.12  &  0.87 \\
  HD76700  &   5737 &  4.25  &  1.18  &  0.41  &  0.20  &  0.40  &  0.23  &  0.36  &  0.39  &  0.71  &  1.00 \\
  HD89744  &   6234 &  3.98  &  1.62  &  0.22  &  0.00  &  0.17  &  0.17  &  0.16  &  0.26  &  0.98  &  0.85 \\
  HD95128  &   5954 &  4.44  &  1.30  &  0.06  &  0.05  &  0.07  &  0.03  &  0.07  &  0.05  &  0.63  &  1.12 \\
 HD106252  &   5899 &  4.34  &  1.08  & -0.01  & -0.16  & -0.07  & -0.10  & -0.02  & -0.05  &  0.76  &  1.15 \\
 HD114762  &   5884 &  4.22  &  1.31  & -0.70  & -0.45  & -0.73  & -0.37  & -0.40  & -0.51  &  0.79  &  1.38 \\
 HD143761  &   5853 &  4.41  &  1.35  & -0.21  & -0.10  & -0.22  & -0.13  & -0.12  & -0.12  &  0.62  &  1.07 \\
 HD145675  &   5311 &  4.42  &  0.92  &  0.43  &  0.37  &  0.45  &  0.40  &  0.36  &  0.42  &  0.71  &  0.93 \\
 HD150706  &   5961 &  4.50  &  1.11  & -0.01  & -0.19  & -0.10  & -0.15  & -0.12  & -0.05  &  0.72  &  0.91 \\
 HD168443  &   5617 &  4.22  &  1.21  &  0.06  &  0.14  &  0.08  &  0.17  &  0.12  &  0.11  &  0.71  &  1.10 \\
HD178911B  &   5600 &  4.44  &  0.95  &  0.27  &  0.13  &  0.28  &  0.44  &  0.19  &  0.16  &  1.35  &  1.15 \\
 HD183263  &   5991 &  4.38  &  1.23  &  0.34  &  0.32  &  0.37  &  0.23  &  0.24  &  0.33  &  0.54  &  0.87 \\
 HD186427  &   5772 &  4.40  &  1.07  &  0.08  & -0.02  &  0.08  &  0.04  &  0.07  &  0.06  &  0.76  &  1.10 \\
 HD187123  &   5845 &  4.42  &  1.10  &  0.13  &  0.16  &  0.13  &  0.06  &  0.09  &  0.13  &  0.52  &  0.98 \\
 HD190228  &   5327 &  3.90  &  1.11  & -0.26  & -0.24  & -0.28  & -0.27  & -0.22  & -0.19  &  0.62  &  1.00 \\
HD190360A  &   5584 &  4.37  &  1.07  &  0.24  &  0.21  &  0.24  &  0.21  &  0.23  &  0.28  &  0.66  &  0.95 \\
HD195019A  &   5842 &  4.32  &  1.27  &  0.09  & -0.13  &  0.03  &  0.10  &  0.05  &  0.08  &  1.12  &  1.00 \\
 HD216437  &   5887 &  4.30  &  1.31  &  0.25  &  0.12  &  0.28  &  0.22  &  0.21  &  0.25  &  0.83  &  0.98 \\
 HD217014  &   5804 &  4.42  &  1.20  &  0.20  &  0.08  &  0.23  &  0.19  &  0.17  &  0.22  &  0.85  &  0.95 \\
 HD217107  &   5646 &  4.31  &  1.06  &  0.37  &  0.32  &  0.39  &  0.31  &  0.34  &  0.34  &  0.65  &  1.07 \\
  HD88133  &   5438 &  3.94  &  1.16  &  0.33  &  0.20  &  0.34  &  0.24  &  0.27  &  0.34  &  0.72  &  0.91 \\
\hline
\end{tabular}
\end{table}

\clearpage

\begin{table}
\caption{Comparison sample stars from HARPS GTO survey.}
\label{lista_comp1}
\begin{tabular}{lcccrrrrrrrr}
\noalign{\medskip} 
\hline
\noalign{\medskip} 
Star & T$_{\rm eff}$ & log \textit{g} & $\xi_{t}$ & [Fe/H] & [O/H] & [Ni/H] & [C/H]& [Mg/H] & [Si/H]& C/O & Mg/Si\\
& K & & km s$^{-1}$ & & & & & & & &\\
\noalign{\medskip} 
\hline
\hline
\noalign{\medskip} 
    HD283  &  5157. &  4.51 &  0.45 & -0.54 & -0.17 & -0.55 & -0.47 & -0.43 & -0.47 &  0.33 &  1.17 \\
    HD361  &  5913. &  4.60 &  1.00 & -0.12 & -0.11 & -0.16 & -0.16 & -0.16 & -0.12 &  0.59 &  0.98 \\
    HD870  &  5381. &  4.42 &  0.79 & -0.10 & -0.12 & -0.14 & -0.15 & -0.09 & -0.12 &  0.62 &  1.15 \\
    HD967  &  5564. &  4.51 &  0.79 & -0.68 & -0.17 & -0.65 & -0.40 & -0.40 & -0.45 &  0.39 &  1.20 \\
   HD1320  &  5679. &  4.49 &  0.85 & -0.27 & -0.24 & -0.31 & -0.25 & -0.26 & -0.26 &  0.65 &  1.07 \\
   HD1388  &  5954. &  4.41 &  1.13 & -0.01 & -0.09 & -0.01 & -0.07 & -0.03 & -0.01 &  0.69 &  1.02 \\
   HD1461  &  5765. &  4.38 &  0.97 &  0.19 &  0.06 &  0.24 &  0.15 &  0.17 &  0.19 &  0.81 &  1.02 \\
   HD1581  &  5977. &  4.51 &  1.12 & -0.18 & -0.20 & -0.20 & -0.20 & -0.14 & -0.15 &  0.66 &  1.10 \\
   HD2071  &  5719. &  4.47 &  0.95 & -0.09 & -0.17 & -0.11 & -0.12 & -0.11 & -0.09 &  0.74 &  1.02 \\
   HD3569  &  5155. &  4.54 &  0.60 & -0.32 & -0.15 & -0.33 & -0.26 & -0.30 & -0.28 &  0.51 &  1.02 \\
   HD3823  &  6022. &  4.31 &  1.39 & -0.28 & -0.14 & -0.29 & -0.21 & -0.19 & -0.22 &  0.56 &  1.15 \\
   HD4307  &  5812. &  4.10 &  1.22 & -0.23 & -0.18 & -0.24 & -0.18 & -0.15 & -0.17 &  0.66 &  1.12 \\
   HD4915  &  5658. &  4.52 &  0.90 & -0.21 & -0.26 & -0.24 & -0.23 & -0.19 & -0.19 &  0.71 &  1.07 \\
   HD6348  &  5107. &  4.51 &  0.07 & -0.56 & -0.35 & -0.59 & -0.28 & -0.43 & -0.48 &  0.78 &  1.20 \\
   HD6735  &  6082. &  4.49 &  1.15 & -0.06 & -0.11 & -0.10 & -0.08 & -0.09 & -0.05 &  0.71 &  0.98 \\
   HD7134  &  5940. &  4.41 &  1.17 & -0.29 & -0.16 & -0.31 & -0.26 & -0.23 & -0.25 &  0.52 &  1.12 \\
   HD7199  &  5386. &  4.34 &  1.01 &  0.28 &  0.17 &  0.37 &  0.28 &  0.32 &  0.34 &  0.85 &  1.02 \\
   HD7449  &  6024. &  4.51 &  1.11 & -0.11 & -0.11 & -0.15 & -0.16 & -0.16 & -0.11 &  0.59 &  0.95 \\
  HD8389A  &  5283. &  4.37 &  1.06 &  0.34 &  0.35 &  0.44 &  0.37 &  0.36 &  0.40 &  0.69 &  0.98 \\
   HD8406  &  5726. &  4.50 &  0.87 & -0.10 & -0.26 & -0.14 & -0.15 & -0.11 & -0.11 &  0.85 &  1.07 \\
   HD8638  &  5507. &  4.43 &  0.74 & -0.38 & -0.17 & -0.34 & -0.19 & -0.16 & -0.21 &  0.63 &  1.20 \\
   HD8828  &  5403. &  4.46 &  0.72 & -0.16 & -0.13 & -0.16 & -0.23 & -0.16 & -0.14 &  0.52 &  1.02 \\
   HD8859  &  5502. &  4.41 &  0.77 & -0.09 & -0.09 & -0.08 & -0.10 & -0.10 & -0.08 &  0.65 &  1.02 \\
   HD8912  &  5211. &  4.43 &  0.70 & -0.07 & -0.21 & -0.08 & -0.07 & -0.10 & -0.05 &  0.91 &  0.95 \\
   HD9782  &  6023. &  4.42 &  1.09 &  0.09 &  0.05 &  0.10 &  0.07 &  0.02 &  0.07 &  0.69 &  0.95 \\
   HD9796  &  5179. &  4.38 &  0.66 & -0.25 & -0.31 & -0.25 & -0.03 & -0.20 & -0.18 &  1.26 &  1.02 \\
  HD10002  &  5313. &  4.40 &  0.82 &  0.17 &  0.09 &  0.20 &  0.22 &  0.16 &  0.20 &  0.89 &  0.98 \\
  HD10166  &  5221. &  4.48 &  0.74 & -0.39 & -0.23 & -0.41 & -0.33 & -0.38 & -0.34 &  0.52 &  0.98 \\
  HD10180  &  5911. &  4.39 &  1.11 &  0.08 &  0.03 &  0.11 &  0.09 &  0.08 &  0.10 &  0.76 &  1.02 \\
  HD10700  &  5310. &  4.44 &  0.55 & -0.52 & -0.31 & -0.50 & -0.28 & -0.31 & -0.37 &  0.71 &  1.23 \\
  HD11226  &  6098. &  4.35 &  1.28 &  0.04 &  0.04 &  0.06 &  0.09 &  0.01 &  0.06 &  0.74 &  0.95 \\
  HD11505  &  5752. &  4.38 &  0.99 & -0.22 &  0.01 & -0.20 & -0.09 & -0.06 & -0.12 &  0.52 &  1.23 \\
  HD12345  &  5395. &  4.44 &  0.69 & -0.21 & -0.06 & -0.21 & -0.15 & -0.18 & -0.17 &  0.54 &  1.05 \\
  HD12387  &  5700. &  4.39 &  0.93 & -0.24 &  0.01 & -0.21 & -0.02 & -0.07 & -0.11 &  0.62 &  1.17 \\
  HD13060  &  5255. &  4.34 &  0.82 &  0.02 & -0.12 &  0.03 & -0.06 &  0.03 &  0.03 &  0.76 &  1.07 \\
  HD13724  &  5868. &  4.52 &  1.02 &  0.23 &  0.12 &  0.26 &  0.12 &  0.18 &  0.22 &  0.66 &  0.98 \\
  HD14374  &  5425. &  4.48 &  0.81 & -0.04 & -0.11 & -0.06 & -0.07 & -0.06 & -0.03 &  0.72 &  1.00 \\
  HD14747  &  5516. &  4.43 &  0.72 & -0.39 & -0.11 & -0.37 & -0.18 & -0.19 & -0.24 &  0.56 &  1.20 \\
  HD15337  &  5179. &  4.39 &  0.70 &  0.06 &  0.21 &  0.09 &  0.09 &  0.06 &  0.09 &  0.50 &  1.00 \\
  HD16297  &  5422. &  4.47 &  0.80 & -0.01 & -0.07 & -0.03 & -0.08 & -0.07 & -0.02 &  0.65 &  0.95 \\
  HD16714  &  5518. &  4.42 &  0.76 & -0.20 & -0.13 & -0.20 & -0.16 & -0.15 & -0.16 &  0.62 &  1.10 \\
  HD18386  &  5457. &  4.39 &  0.92 &  0.14 &  0.00 &  0.19 &  0.16 &  0.09 &  0.18 &  0.95 &  0.87 \\
  HD18719  &  5241. &  4.41 &  0.92 & -0.08 & -0.03 & -0.10 & -0.04 & -0.14 & -0.04 &  0.65 &  0.85 \\
  HD19034  &  5477. &  4.40 &  0.69 & -0.48 & -0.09 & -0.45 & -0.21 & -0.18 & -0.29 &  0.50 &  1.38 \\
\hline
\end{tabular}
\end{table}

\begin{table}
\caption{Comparison sample stars from HARPS GTO survey.}
\label{lista_comp2}
\begin{tabular}{lcccrrrrrrrr}
\noalign{\medskip} 
\hline
\noalign{\medskip} 
Star & T$_{\rm eff}$ & log \textit{g} & $\xi_{t}$ & [Fe/H] & [O/H] & [Ni/H] & [C/H]& [Mg/H] & [Si/H]& C/O & Mg/Si\\
& K & & km s$^{-1}$ & & & & & & & &\\
\noalign{\medskip} 
\hline
\hline
\noalign{\medskip} 

  HD19467  &  5720. &  4.31 &  0.96 & -0.14 &  0.01 & -0.12 &  0.02 &  0.01 & -0.06 &  0.68 &  1.26 \\
  HD20003  &  5494. &  4.41 &  0.83 &  0.04 & -0.02 &  0.05 &  0.02 &  0.01 &  0.04 &  0.72 &  1.00 \\
  HD20407  &  5866. &  4.50 &  1.09 & -0.44 & -0.29 & -0.44 & -0.31 & -0.38 & -0.36 &  0.63 &  1.02 \\
  HD20619  &  5703. &  4.51 &  0.92 & -0.22 & -0.27 & -0.26 & -0.25 & -0.22 & -0.21 &  0.69 &  1.05 \\
  HD20781  &  5256. &  4.37 &  0.78 & -0.11 & -0.12 & -0.13 & -0.10 & -0.10 & -0.09 &  0.69 &  1.05 \\
  HD20794  &  5401. &  4.40 &  0.67 & -0.40 & -0.17 & -0.35 & -0.14 & -0.14 & -0.21 &  0.71 &  1.26 \\
  HD20807  &  5866. &  4.52 &  1.04 & -0.23 & -0.18 & -0.24 & -0.19 & -0.20 & -0.20 &  0.65 &  1.07 \\
  HD21019  &  5468. &  3.93 &  1.05 & -0.45 & -0.23 & -0.45 & -0.40 & -0.31 & -0.37 &  0.45 &  1.23 \\
  HD21411  &  5473. &  4.51 &  0.81 & -0.26 & -0.15 & -0.30 & -0.29 & -0.26 & -0.26 &  0.48 &  1.07 \\
  HD21693  &  5430. &  4.37 &  0.76 &  0.00 & -0.15 &  0.03 &  0.01 &  0.01 &  0.03 &  0.95 &  1.02 \\
  HD23249  &  5150. &  3.89 &  1.01 &  0.13 &  0.22 &  0.16 &  0.14 &  0.19 &  0.16 &  0.55 &  1.15 \\
  HD23456  &  6178. &  4.56 &  1.38 & -0.32 & -0.19 & -0.34 & -0.24 & -0.30 & -0.26 &  0.59 &  0.98 \\
  HD24892  &  5363. &  3.99 &  0.88 & -0.32 & -0.10 & -0.31 & -0.20 & -0.17 & -0.21 &  0.52 &  1.17 \\
  HD25105  &  5316. &  4.47 &  0.77 & -0.15 & -0.14 & -0.17 & -0.18 & -0.17 & -0.15 &  0.60 &  1.02 \\
  HD25120  &  5134. &  4.47 &  0.87 & -0.18 & -0.17 & -0.24 & -0.09 & -0.21 & -0.18 &  0.79 &  1.00 \\
  HD25565  &  5212. &  4.47 &  0.80 &  0.03 &  0.02 &  0.03 &  0.02 & -0.01 &  0.04 &  0.66 &  0.95 \\
  HD25673  &  5136. &  4.47 &  0.56 & -0.50 & -0.35 & -0.50 & -0.18 & -0.44 & -0.43 &  0.98 &  1.05 \\
 HD26965A  &  5153. &  4.39 &  0.36 & -0.31 & -0.14 & -0.26 & -0.12 & -0.10 & -0.18 &  0.69 &  1.29 \\
  HD27063  &  5767. &  4.44 &  0.94 &  0.05 & -0.03 &  0.04 & -0.04 &  0.01 &  0.03 &  0.65 &  1.02 \\
  HD28471  &  5745. &  4.37 &  0.95 & -0.05 & -0.14 & -0.06 & -0.07 & -0.04 & -0.04 &  0.78 &  1.07 \\
  HD28701  &  5710. &  4.41 &  0.95 & -0.32 & -0.12 & -0.29 & -0.12 & -0.13 & -0.16 &  0.66 &  1.15 \\
  HD28821  &  5660. &  4.38 &  0.88 & -0.12 & -0.19 & -0.15 & -0.10 & -0.06 & -0.09 &  0.81 &  1.15 \\
  HD30278  &  5394. &  4.39 &  0.72 & -0.17 & -0.05 & -0.20 & -0.15 & -0.12 & -0.15 &  0.52 &  1.15 \\
  HD30306  &  5529. &  4.32 &  0.89 &  0.17 &  0.06 &  0.18 &  0.14 &  0.18 &  0.19 &  0.79 &  1.05 \\
  HD31527  &  5898. &  4.45 &  1.09 & -0.17 & -0.22 & -0.18 & -0.17 & -0.16 & -0.16 &  0.74 &  1.07 \\
  HD31822  &  6042. &  4.57 &  1.15 & -0.19 & -0.18 & -0.24 & -0.25 & -0.19 & -0.19 &  0.56 &  1.07 \\
  HD32724  &  5818. &  4.26 &  1.14 & -0.17 & -0.08 & -0.19 & -0.14 & -0.09 & -0.12 &  0.58 &  1.15 \\
  HD33725  &  5274. &  4.41 &  0.71 & -0.17 & -0.09 & -0.16 & -0.16 & -0.16 & -0.15 &  0.56 &  1.05 \\
  HD34449  &  5848. &  4.50 &  0.92 & -0.09 & -0.08 & -0.13 & -0.13 & -0.12 & -0.11 &  0.59 &  1.05 \\
  HD34688  &  5169. &  4.44 &  0.70 & -0.20 & -0.20 & -0.20 & -0.10 & -0.19 & -0.19 &  0.83 &  1.07 \\
  HD36108  &  5916. &  4.33 &  1.21 & -0.21 & -0.14 & -0.25 & -0.19 & -0.17 & -0.19 &  0.59 &  1.12 \\
  HD36379  &  6030. &  4.30 &  1.29 & -0.17 & -0.14 & -0.18 & -0.11 & -0.12 & -0.13 &  0.71 &  1.10 \\
  HD37962  &  5718. &  4.48 &  0.84 & -0.20 & -0.12 & -0.25 & -0.25 & -0.23 & -0.19 &  0.49 &  0.98 \\
  HD37986  &  5507. &  4.29 &  0.92 &  0.26 &  0.24 &  0.32 &  0.28 &  0.25 &  0.30 &  0.72 &  0.95 \\
  HD38277  &  5871. &  4.34 &  1.10 & -0.07 & -0.07 & -0.09 & -0.06 & -0.04 & -0.07 &  0.68 &  1.15 \\
  HD38382  &  6082. &  4.45 &  1.18 &  0.03 & -0.18 &  0.00 & -0.01 &  0.00 &  0.02 &  0.98 &  1.02 \\
  HD38858  &  5733. &  4.51 &  0.94 & -0.22 & -0.23 & -0.25 & -0.21 & -0.23 & -0.21 &  0.69 &  1.02 \\
  HD38973  &  6016. &  4.42 &  1.14 &  0.05 & -0.07 &  0.06 &  0.05 &  0.01 &  0.05 &  0.87 &  0.98 \\
  HD39194  &  5205. &  4.53 &  0.37 & -0.61 & -0.23 & -0.57 & -0.28 & -0.42 & -0.43 &  0.59 &  1.10 \\
  HD40105  &  5137. &  3.85 &  0.97 &  0.06 &  0.11 &  0.09 &  0.06 &  0.09 &  0.10 &  0.59 &  1.05 \\
  HD40397  &  5527. &  4.39 &  0.83 & -0.13 &  0.10 & -0.11 &  0.00 & -0.02 & -0.06 &  0.52 &  1.17 \\
  HD44120  &  6052. &  4.25 &  1.31 &  0.12 &  0.05 &  0.14 &  0.08 &  0.10 &  0.13 &  0.71 &  1.00 \\
  HD44420  &  5818. &  4.37 &  1.06 &  0.29 &  0.15 &  0.38 &  0.29 &  0.29 &  0.33 &  0.91 &  0.98 \\
  HD44447  &  5999. &  4.37 &  1.26 & -0.22 & -0.11 & -0.23 & -0.16 & -0.20 & -0.18 &  0.59 &  1.02 \\
  HD44594  &  5840. &  4.38 &  1.06 &  0.15 & -0.02 &  0.18 &  0.10 &  0.13 &  0.15 &  0.87 &  1.02 \\
\hline
\end{tabular}
\end{table}

\begin{table}
\caption{Comparison sample stars from HARPS GTO survey.}
\label{lista_comp3}
\begin{tabular}{lcccrrrrrrrr}
\noalign{\medskip} 
\hline
\noalign{\medskip} 
Star & T$_{\rm eff}$ & log \textit{g} & $\xi_{t}$ & [Fe/H] & [O/H] & [Ni/H] & [C/H]& [Mg/H] & [Si/H]& C/O & Mg/Si\\
& K & & km s$^{-1}$ & & & & & & & &\\
\noalign{\medskip} 
\hline
\hline
\noalign{\medskip} 

  HD45184  &  5869. &  4.47 &  1.03 &  0.04 & -0.02 &  0.03 & -0.02 &  0.03 &  0.04 &  0.66 &  1.05 \\
  HD45289  &  5717. &  4.32 &  0.99 & -0.02 &  0.10 & -0.01 &  0.03 &  0.06 &  0.03 &  0.56 &  1.15 \\
  HD48611  &  5337. &  4.51 &  0.69 & -0.36 & -0.31 & -0.37 & -0.31 & -0.35 & -0.31 &  0.66 &  0.98 \\
  HD50806  &  5633. &  4.11 &  1.03 &  0.03 &  0.29 &  0.07 &  0.14 &  0.15 &  0.10 &  0.47 &  1.20 \\
  HD51608  &  5358. &  4.36 &  0.73 & -0.07 & -0.11 & -0.07 & -0.07 & -0.02 & -0.02 &  0.72 &  1.07 \\
  HD55693  &  5914. &  4.43 &  1.07 &  0.29 &  0.12 &  0.36 &  0.29 &  0.26 &  0.31 &  0.98 &  0.95 \\
  HD59468  &  5618. &  4.39 &  0.88 &  0.03 & -0.01 &  0.05 &  0.05 &  0.06 &  0.03 &  0.76 &  1.15 \\
 HD59711A  &  5722. &  4.46 &  0.86 & -0.12 & -0.14 & -0.13 & -0.14 &  0.06 &  0.03 &  0.66 &  1.15 \\
  HD63765  &  5432. &  4.42 &  0.82 & -0.16 & -0.21 & -0.19 & -0.22 & -0.17 & -0.15 &  0.65 &  1.02 \\
 HD65907A  &  5945. &  4.52 &  1.05 & -0.31 &  0.00 & -0.27 & -0.08 & -0.05 & -0.13 &  0.55 &  1.29 \\
  HD66221  &  5635. &  4.40 &  0.92 &  0.17 &  0.03 &  0.22 &  0.14 &  0.17 &  0.18 &  0.85 &  1.05 \\
  HD67458  &  5891. &  4.53 &  1.04 & -0.16 & -0.18 & -0.18 & -0.21 & -0.16 & -0.15 &  0.62 &  1.05 \\
  HD68607  &  5215. &  4.41 &  0.82 &  0.07 & -0.04 &  0.08 &  0.08 &  0.02 &  0.09 &  0.87 &  0.91 \\
 HD68978A  &  5965. &  4.48 &  1.09 &  0.04 & -0.08 &  0.04 &  0.00 &  0.00 &  0.04 &  0.79 &  0.98 \\
  HD69655  &  5961. &  4.44 &  1.15 & -0.18 & -0.29 & -0.19 & -0.16 & -0.16 & -0.17 &  0.89 &  1.10 \\
  HD70889  &  6051. &  4.49 &  1.13 &  0.11 &  0.00 &  0.09 & -0.01 &  0.01 &  0.08 &  0.65 &  0.91 \\
  HD71334  &  5694. &  4.37 &  0.95 & -0.09 & -0.11 & -0.10 & -0.12 & -0.04 & -0.08 &  0.65 &  1.17 \\
  HD71479  &  6026. &  4.42 &  1.19 &  0.24 &  0.33 &  0.29 &  0.19 &  0.20 &  0.25 &  0.48 &  0.95 \\
  HD71835  &  5438. &  4.39 &  0.79 & -0.04 & -0.17 & -0.03 & -0.04 & -0.06 & -0.02 &  0.89 &  0.98 \\
  HD72579  &  5449. &  4.27 &  0.84 &  0.20 &  0.12 &  0.22 &  0.21 &  0.21 &  0.22 &  0.81 &  1.05 \\
  HD72673  &  5243. &  4.46 &  0.60 & -0.41 & -0.33 & -0.41 & -0.25 & -0.34 & -0.36 &  0.79 &  1.12 \\
  HD72769  &  5640. &  4.35 &  0.98 &  0.30 &  0.20 &  0.36 &  0.27 &  0.27 &  0.30 &  0.78 &  1.00 \\
  HD73121  &  6091. &  4.30 &  1.34 &  0.09 &  0.01 &  0.10 &  0.06 &  0.07 &  0.12 &  0.74 &  0.95 \\
  HD73524  &  6017. &  4.43 &  1.14 &  0.16 &  0.02 &  0.16 &  0.00 &  0.11 &  0.15 &  0.63 &  0.98 \\
  HD74014  &  5561. &  4.33 &  0.90 &  0.22 &  0.05 &  0.27 &  0.23 &  0.20 &  0.26 &  1.00 &  0.93 \\
  HD76151  &  5788. &  4.48 &  0.96 &  0.12 &  0.04 &  0.14 &  0.07 &  0.10 &  0.11 &  0.71 &  1.05 \\
  HD78429  &  5760. &  4.33 &  1.01 &  0.09 &  0.27 &  0.07 &  0.08 &  0.08 &  0.09 &  0.43 &  1.05 \\
  HD78538  &  5786. &  4.50 &  0.98 & -0.03 & -0.06 & -0.08 & -0.19 & -0.08 & -0.05 &  0.49 &  1.00 \\
  HD78558  &  5711. &  4.36 &  0.99 & -0.44 & -0.08 & -0.41 & -0.15 & -0.16 & -0.23 &  0.56 &  1.26 \\
  HD78747  &  5778. &  4.46 &  1.03 & -0.67 & -0.21 & -0.65 & -0.38 & -0.38 & -0.44 &  0.45 &  1.23 \\
  HD80883  &  5233. &  4.44 &  0.80 & -0.25 & -0.32 & -0.28 & -0.29 & -0.25 & -0.22 &  0.71 &  1.00 \\
  HD81639  &  5522. &  4.40 &  0.79 & -0.17 & -0.22 & -0.16 & -0.25 & -0.11 & -0.14 &  0.62 &  1.15 \\
  HD82516  &  5104. &  4.46 &  0.71 &  0.01 &  0.07 &  0.04 &  0.18 & -0.02 &  0.02 &  0.85 &  0.98 \\
  HD83529  &  5902. &  4.35 &  1.11 & -0.22 & -0.20 & -0.25 & -0.24 & -0.16 & -0.20 &  0.60 &  1.17 \\
  HD85119  &  5425. &  4.52 &  0.93 & -0.20 & -0.17 & -0.26 & -0.21 & -0.24 & -0.20 &  0.60 &  0.98 \\
  HD85390  &  5186. &  4.41 &  0.75 & -0.07 & -0.13 & -0.06 &  0.04 & -0.08 & -0.03 &  0.98 &  0.95 \\
  HD86171  &  5400. &  4.47 &  0.81 & -0.25 & -0.34 & -0.29 & -0.32 & -0.24 & -0.22 &  0.69 &  1.02 \\
  HD88218  &  5878. &  4.16 &  1.23 & -0.14 & -0.12 & -0.15 & -0.08 & -0.10 & -0.11 &  0.72 &  1.10 \\
  HD88656  &  5150. &  4.44 &  0.81 & -0.11 & -0.21 & -0.14 & -0.18 & -0.16 & -0.10 &  0.71 &  0.93 \\
  HD88742  &  5981. &  4.52 &  1.07 & -0.02 & -0.10 & -0.05 & -0.04 & -0.08 & -0.03 &  0.76 &  0.95 \\
  HD89454  &  5728. &  4.47 &  0.96 &  0.12 & -0.03 &  0.12 &  0.01 &  0.07 &  0.10 &  0.72 &  1.00 \\
  HD90156  &  5599. &  4.48 &  0.86 & -0.24 & -0.24 & -0.25 & -0.30 & -0.18 & -0.21 &  0.58 &  1.15 \\
  HD90711  &  5444. &  4.40 &  0.92 &  0.24 &  0.16 &  0.31 &  0.23 &  0.24 &  0.27 &  0.78 &  1.00 \\
  HD90812  &  5164. &  4.48 &  0.64 & -0.36 & -0.28 & -0.35 & -0.28 & -0.30 & -0.30 &  0.66 &  1.07 \\
  HD92588  &  5199. &  3.79 &  1.01 &  0.04 &  0.08 &  0.04 & -0.05 &  0.05 &  0.05 &  0.49 &  1.07 \\

\hline
\end{tabular}
\end{table}

\begin{table}
\caption{Comparison sample stars from HARPS GTO survey.}
\label{lista_comp4}
\begin{tabular}{lcccrrrrrrrr}
\noalign{\medskip} 
\hline
\noalign{\medskip} 
Star & T$_{\rm eff}$ & log \textit{g} & $\xi_{t}$ & [Fe/H] & [O/H] & [Ni/H] & [C/H]& [Mg/H] & [Si/H]& C/O & Mg/Si\\
& K & & km s$^{-1}$ & & & & & & & &\\
\noalign{\medskip} 
\hline
\hline
\noalign{\medskip} 

  HD92719  &  5824. &  4.51 &  0.96 & -0.10 & -0.16 & -0.13 & -0.18 & -0.12 & -0.10 &  0.63 &  1.02 \\
  HD93385  &  5977. &  4.42 &  1.14 &  0.02 & -0.03 &  0.03 & -0.01 &  0.00 &  0.02 &  0.69 &  1.02 \\
  HD94151  &  5583. &  4.38 &  0.83 &  0.04 & -0.10 &  0.06 &  0.03 &  0.03 &  0.05 &  0.89 &  1.02 \\
  HD95456  &  6276. &  4.35 &  1.40 &  0.16 &  0.10 &  0.15 &  0.09 &  0.10 &  0.15 &  0.65 &  0.95 \\
  HD95521  &  5773. &  4.49 &  0.96 & -0.15 & -0.16 & -0.18 & -0.17 & -0.17 & -0.15 &  0.65 &  1.02 \\
  HD96423  &  5711. &  4.35 &  0.98 &  0.10 & -0.01 &  0.13 &  0.07 &  0.12 &  0.11 &  0.79 &  1.10 \\
  HD96700  &  5845. &  4.39 &  1.04 & -0.18 & -0.10 & -0.21 & -0.15 & -0.15 & -0.16 &  0.59 &  1.10 \\
  HD97037  &  5883. &  4.34 &  1.13 & -0.07 & -0.10 & -0.09 & -0.08 & -0.06 & -0.05 &  0.69 &  1.05 \\
  HD97343  &  5410. &  4.39 &  0.82 & -0.06 & -0.04 & -0.06 & -0.06 &  0.03 & -0.01 &  0.63 &  1.17 \\
  HD97998  &  5716. &  4.57 &  0.85 & -0.42 & -0.24 & -0.44 & -0.38 & -0.36 & -0.37 &  0.48 &  1.10 \\
  HD98281  &  5381. &  4.42 &  0.64 & -0.26 & -0.25 & -0.26 & -0.26 & -0.19 & -0.21 &  0.65 &  1.12 \\
  HD98356  &  5322. &  4.41 &  0.84 &  0.10 &  0.03 &  0.13 &  0.09 &  0.08 &  0.12 &  0.76 &  0.98 \\
 HD100508  &  5449. &  4.42 &  0.86 &  0.39 &  0.25 &  0.49 &  0.31 &  0.37 &  0.41 &  0.76 &  0.98 \\
 HD102365  &  5629. &  4.44 &  0.91 & -0.29 & -0.11 & -0.30 & -0.17 & -0.23 & -0.19 &  0.58 &  0.98 \\
 HD102438  &  5560. &  4.41 &  0.84 & -0.29 & -0.26 & -0.30 & -0.24 & -0.22 & -0.24 &  0.69 &  1.12 \\
 HD104263  &  5477. &  4.34 &  0.81 &  0.02 &  0.01 &  0.04 &  0.09 &  0.08 &  0.06 &  0.79 &  1.12 \\
 HD104982  &  5692. &  4.44 &  0.91 & -0.19 & -0.24 & -0.20 & -0.18 & -0.15 & -0.17 &  0.76 &  1.12 \\
 HD105837  &  5907. &  4.54 &  1.14 & -0.51 & -0.37 & -0.51 & -0.13 & -0.45 & -0.45 &  1.15 &  1.07 \\
 HD106116  &  5680. &  4.39 &  0.91 &  0.14 &  0.12 &  0.17 &  0.09 &  0.11 &  0.14 &  0.62 &  1.00 \\
 HD108309  &  5775. &  4.23 &  1.08 &  0.12 &  0.10 &  0.13 &  0.15 &  0.13 &  0.13 &  0.74 &  1.07 \\
 HD109200  &  5134. &  4.51 &  0.68 & -0.31 & -0.20 & -0.33 & -0.19 & -0.27 & -0.28 &  0.68 &  1.10 \\
 HD109409  &  5886. &  4.16 &  1.24 &  0.33 &  0.22 &  0.40 &  0.24 &  0.24 &  0.32 &  0.69 &  0.89 \\
 HD111031  &  5801. &  4.39 &  1.05 &  0.27 &  0.21 &  0.34 &  0.24 &  0.23 &  0.27 &  0.71 &  0.98 \\
 HD112540  &  5523. &  4.52 &  0.74 & -0.17 & -0.26 & -0.20 & -0.22 & -0.20 & -0.17 &  0.72 &  1.00 \\
 HD114613  &  5729. &  3.97 &  1.18 &  0.19 &  0.10 &  0.24 &  0.13 &  0.17 &  0.21 &  0.71 &  0.98 \\
 HD114747  &  5172. &  4.44 &  0.98 &  0.21 &  0.18 &  0.30 &  0.35 &  0.21 &  0.28 &  0.98 &  0.91 \\
 HD114853  &  5705. &  4.44 &  0.92 & -0.23 &  0.01 & -0.25 & -0.25 & -0.19 & -0.20 &  0.36 &  1.10 \\
 HD115585  &  5711. &  4.27 &  1.14 &  0.35 &  0.22 &  0.40 &  0.34 &  0.33 &  0.35 &  0.87 &  1.02 \\
 HD115617  &  5558. &  4.36 &  0.81 & -0.02 & -0.08 & -0.01 & -0.11 &  0.00 &  0.00 &  0.62 &  1.07 \\
 HD115674  &  5649. &  4.48 &  0.85 & -0.17 & -0.28 & -0.20 & -0.22 & -0.22 & -0.17 &  0.76 &  0.95 \\
 HD117105  &  5889. &  4.41 &  1.13 & -0.29 & -0.28 & -0.32 & -0.23 & -0.19 & -0.22 &  0.74 &  1.15 \\
 HD119638  &  6069. &  4.42 &  1.22 & -0.15 & -0.11 & -0.17 & -0.13 & -0.16 & -0.12 &  0.63 &  0.98 \\
 HD119782  &  5160. &  4.44 &  0.79 & -0.07 & -0.10 & -0.07 &  0.00 & -0.10 & -0.05 &  0.83 &  0.95 \\
 HD122862  &  5982. &  4.23 &  1.29 & -0.12 & -0.09 & -0.12 & -0.08 & -0.12 & -0.09 &  0.68 &  1.00 \\
 HD123265  &  5338. &  4.29 &  0.85 &  0.19 &  0.14 &  0.22 &  0.27 &  0.27 &  0.27 &  0.89 &  1.07 \\
 HD124106  &  5106. &  4.49 &  0.80 & -0.17 & -0.17 & -0.20 & -0.18 & -0.23 & -0.14 &  0.65 &  0.87 \\
 HD124292  &  5443. &  4.37 &  0.77 & -0.13 & -0.12 & -0.12 & -0.07 & -0.07 & -0.09 &  0.74 &  1.12 \\
 HD124364  &  5584. &  4.48 &  0.83 & -0.27 & -0.28 & -0.31 & -0.31 & -0.26 & -0.26 &  0.62 &  1.07 \\
 HD125184  &  5680. &  4.10 &  1.13 &  0.27 &  0.15 &  0.28 &  0.18 &  0.25 &  0.27 &  0.71 &  1.02 \\
 HD125455  &  5162. &  4.52 &  0.70 & -0.18 & -0.18 & -0.18 & -0.06 & -0.18 & -0.17 &  0.87 &  1.05 \\
 HD125881  &  6036. &  4.49 &  1.10 &  0.06 & -0.05 &  0.05 & -0.01 &  0.00 &  0.05 &  0.72 &  0.95 \\
 HD126525  &  5638. &  4.37 &  0.90 & -0.10 & -0.11 & -0.08 & -0.07 & -0.08 & -0.08 &  0.72 &  1.07 \\
 HD128674  &  5551. &  4.50 &  0.71 & -0.38 & -0.35 & -0.38 & -0.34 & -0.33 & -0.32 &  0.68 &  1.05 \\
 HD132648  &  5418. &  4.49 &  0.69 & -0.37 & -0.38 & -0.38 & -0.35 & -0.33 & -0.34 &  0.71 &  1.10 \\
 HD134060  &  5966. &  4.43 &  1.10 &  0.14 & -0.04 &  0.15 &  0.07 &  0.11 &  0.12 &  0.85 &  1.05 \\
\hline
\end{tabular}
\end{table}

\begin{table}
\caption{Comparison sample stars from HARPS GTO survey.}
\label{lista_comp5}
\begin{tabular}{lcccrrrrrrrr}
\noalign{\medskip} 
\hline
\noalign{\medskip} 
Star & T$_{\rm eff}$ & log \textit{g} & $\xi_{t}$ & [Fe/H] & [O/H] & [Ni/H] & [C/H]& [Mg/H] & [Si/H]& C/O & Mg/Si\\
& K & & km s$^{-1}$ & & & & & & & &\\
\noalign{\medskip} 
\hline
\hline
\noalign{\medskip} 

 HD134606  &  5633. &  4.38 &  1.00 &  0.27 &  0.21 &  0.33 &  0.27 &  0.26 &  0.30 &  0.76 &  0.98 \\
 HD134664  &  5865. &  4.52 &  0.99 &  0.10 & -0.05 &  0.09 & -0.02 &  0.03 &  0.08 &  0.71 &  0.95 \\
 HD136894  &  5412. &  4.36 &  0.75 & -0.10 & -0.11 & -0.13 & -0.12 & -0.11 & -0.11 &  0.65 &  1.07 \\
 HD137388  &  5240. &  4.42 &  0.93 &  0.18 &  0.26 &  0.25 &  0.28 &  0.20 &  0.24 &  0.69 &  0.98 \\
 HD138549  &  5582. &  4.44 &  0.87 &  0.00 & -0.10 &  0.03 &  0.02 & -0.02 &  0.01 &  0.87 &  1.00 \\
 HD140901  &  5610. &  4.46 &  0.90 &  0.09 & -0.04 &  0.11 &  0.05 &  0.05 &  0.08 &  0.81 &  1.00 \\
 HD143114  &  5775. &  4.39 &  0.92 & -0.41 & -0.12 & -0.39 & -0.24 & -0.22 & -0.26 &  0.50 &  1.17 \\
 HD144585  &  5914. &  4.35 &  1.15 &  0.33 &  0.22 &  0.39 &  0.20 &  0.30 &  0.33 &  0.63 &  1.00 \\
 HD145598  &  5417. &  4.48 &  0.59 & -0.78 & -0.44 & -0.75 & -0.52 & -0.46 & -0.54 &  0.55 &  1.29 \\
 HD145666  &  5958. &  4.53 &  1.04 & -0.04 & -0.21 & -0.08 & -0.11 & -0.09 & -0.06 &  0.83 &  1.00 \\
 HD145809  &  5778. &  4.15 &  1.14 & -0.25 & -0.05 & -0.29 & -0.20 & -0.16 & -0.21 &  0.47 &  1.20 \\
 HD146233  &  5818. &  4.45 &  1.00 &  0.04 & -0.04 &  0.04 & -0.03 &  0.03 &  0.04 &  0.68 &  1.05 \\
 HD147512  &  5530. &  4.40 &  0.81 & -0.08 & -0.09 & -0.06 & -0.04 & -0.04 & -0.06 &  0.74 &  1.12 \\
 HD151504  &  5457. &  4.36 &  0.87 &  0.06 &  0.16 &  0.07 &  0.14 &  0.10 &  0.10 &  0.63 &  1.07 \\
 HD154088  &  5374. &  4.37 &  0.85 &  0.28 &  0.19 &  0.35 &  0.29 &  0.25 &  0.30 &  0.83 &  0.95 \\
 HD154962  &  5827. &  4.17 &  1.22 &  0.32 &  0.25 &  0.39 &  0.28 &  0.30 &  0.32 &  0.71 &  1.02 \\
 HD157172  &  5451. &  4.39 &  0.77 &  0.11 & -0.05 &  0.15 &  0.11 &  0.10 &  0.12 &  0.95 &  1.02 \\
 HD157347  &  5676. &  4.38 &  0.91 &  0.02 & -0.10 &  0.03 & -0.03 &  0.00 &  0.02 &  0.78 &  1.02 \\
 HD161098  &  5560. &  4.46 &  0.79 & -0.27 & -0.23 & -0.29 & -0.29 & -0.26 & -0.25 &  0.58 &  1.05 \\
 HD161612  &  5616. &  4.45 &  0.88 &  0.16 &  0.05 &  0.17 &  0.10 &  0.14 &  0.16 &  0.74 &  1.02 \\
 HD162236  &  5343. &  4.43 &  0.82 & -0.12 & -0.13 & -0.13 & -0.14 & -0.15 & -0.10 &  0.65 &  0.95 \\
 HD162396  &  6090. &  4.27 &  1.43 & -0.35 & -0.23 & -0.33 & -0.26 & -0.25 & -0.27 &  0.62 &  1.12 \\
 HD165920  &  5339. &  4.39 &  0.79 &  0.29 &  0.44 &  0.32 &  0.49 &  0.23 &  0.31 &  0.74 &  0.89 \\
 HD166724  &  5127. &  4.43 &  0.79 & -0.09 & -0.20 & -0.10 &  0.01 & -0.12 & -0.10 &  1.07 &  1.02 \\
 HD167359  &  5348. &  4.46 &  0.67 & -0.19 & -0.17 & -0.22 & -0.26 & -0.19 & -0.17 &  0.54 &  1.02 \\
 HD168871  &  5983. &  4.42 &  1.17 & -0.09 & -0.15 & -0.10 & -0.07 & -0.09 & -0.07 &  0.79 &  1.02 \\
 HD171665  &  5655. &  4.41 &  0.89 & -0.05 & -0.14 & -0.05 & -0.09 & -0.07 & -0.03 &  0.74 &  0.98 \\
 HD171990  &  6045. &  4.14 &  1.40 &  0.06 &  0.09 &  0.08 &  0.08 &  0.06 &  0.08 &  0.65 &  1.02 \\
 HD172513  &  5500. &  4.41 &  0.79 & -0.05 & -0.22 & -0.07 & -0.12 & -0.07 & -0.04 &  0.83 &  1.00 \\
 HD174545  &  5216. &  4.40 &  0.88 &  0.22 &  0.12 &  0.28 &  0.30 &  0.21 &  0.25 &  1.00 &  0.98 \\
 HD176157  &  5181. &  4.41 &  0.92 & -0.16 & -0.33 & -0.16 & -0.06 & -0.13 & -0.13 &  1.23 &  1.07 \\
 HD177409  &  5898. &  4.49 &  0.99 & -0.04 & -0.16 & -0.07 & -0.13 & -0.05 & -0.04 &  0.71 &  1.05 \\
 HD177565  &  5627. &  4.39 &  0.91 &  0.08 &  0.16 &  0.11 &  0.06 &  0.06 &  0.09 &  0.52 &  1.00 \\
 HD177758  &  5862. &  4.41 &  1.11 & -0.58 & -0.26 & -0.56 & -0.34 & -0.37 & -0.39 &  0.55 &  1.12 \\
 HD180409  &  6013. &  4.52 &  1.16 & -0.17 & -0.17 & -0.19 & -0.15 & -0.17 & -0.16 &  0.69 &  1.05 \\
 HD183658  &  5803. &  4.40 &  1.00 &  0.03 & -0.32 &  0.05 &  0.03 &  0.01 &  0.03 &  1.48 &  1.02 \\
 HD185615  &  5570. &  4.34 &  0.84 &  0.08 &  0.12 &  0.12 &  0.16 &  0.11 &  0.12 &  0.72 &  1.05 \\
 HD189567  &  5726. &  4.41 &  0.95 & -0.24 & -0.18 & -0.26 & -0.16 & -0.15 & -0.20 &  0.69 &  1.20 \\
 HD189625  &  5846. &  4.43 &  1.03 &  0.18 &  0.31 &  0.20 &  0.08 &  0.13 &  0.19 &  0.39 &  0.93 \\
 HD190248  &  5604. &  4.26 &  0.99 &  0.33 &  0.30 &  0.37 &  0.30 &  0.30 &  0.35 &  0.66 &  0.95 \\
 HD192031  &  5215. &  4.39 &  0.04 & -0.84 & -0.48 & -0.85 & -0.50 & -0.57 & -0.59 &  0.63 &  1.12 \\
 HD192117  &  5479. &  4.48 &  0.75 & -0.04 & -0.31 & -0.07 & -0.02 & -0.09 & -0.07 &  1.29 &  1.02 \\
 HD192310  &  5166. &  4.51 &  0.97 & -0.04 & -0.08 & -0.01 &  0.14 & -0.01 & -0.03 &  1.10 &  1.12 \\
 HD193193  &  5979. &  4.40 &  1.15 & -0.05 &  0.04 & -0.05 & -0.03 & -0.05 & -0.04 &  0.56 &  1.05 \\
 HD195564  &  5676. &  4.03 &  1.11 &  0.06 &  0.02 &  0.05 &  0.04 &  0.09 &  0.06 &  0.69 &  1.15 \\
\hline
\end{tabular}
\end{table}

\begin{table}
\caption{Comparison sample stars from HARPS GTO survey.}
\label{lista_comp6}
\begin{tabular}{lcccrrrrrrrr}
\noalign{\medskip} 
\hline
\noalign{\medskip} 
Star & T$_{\rm eff}$ & log \textit{g} & $\xi_{t}$ & [Fe/H] & [O/H] & [Ni/H] & [C/H]& [Mg/H] & [Si/H]& C/O & Mg/Si\\
& K & & km s$^{-1}$ & & & & & & & &\\
\noalign{\medskip} 
\hline
\hline
\noalign{\medskip} 

 HD196761  &  5415. &  4.43 &  0.76 & -0.31 & -0.39 & -0.31 & -0.26 & -0.29 & -0.27 &  0.89 &  1.02 \\
 HD196800  &  6010. &  4.37 &  1.17 &  0.19 &  0.19 &  0.22 &  0.16 &  0.14 &  0.17 &  0.62 &  1.00 \\
 HD197210  &  5577. &  4.42 &  0.86 & -0.03 & -0.17 & -0.05 & -0.14 & -0.06 & -0.03 &  0.71 &  1.00 \\
 HD197823  &  5396. &  4.41 &  0.82 &  0.12 & -0.06 &  0.15 &  0.11 &  0.08 &  0.14 &  0.98 &  0.93 \\
 HD198075  &  5846. &  4.56 &  0.95 & -0.24 & -0.10 & -0.28 & -0.31 & -0.25 & -0.25 &  0.41 &  1.07 \\
 HD199288  &  5765. &  4.50 &  1.00 & -0.63 & -0.29 & -0.63 & -0.39 & -0.43 & -0.46 &  0.52 &  1.15 \\
 HD199960  &  5973. &  4.39 &  1.13 &  0.28 &  0.23 &  0.34 &  0.23 &  0.24 &  0.29 &  0.66 &  0.95 \\
 HD202605  &  5658. &  4.49 &  1.02 &  0.18 &  0.13 &  0.19 &  0.11 &  0.12 &  0.17 &  0.63 &  0.95 \\
 HD203384  &  5586. &  4.40 &  0.90 &  0.26 &  0.19 &  0.31 &  0.23 &  0.24 &  0.26 &  0.72 &  1.02 \\
 HD203432  &  5645. &  4.39 &  0.98 &  0.29 &  0.16 &  0.36 &  0.22 &  0.29 &  0.29 &  0.76 &  1.07 \\
 HD204385  &  6033. &  4.44 &  1.15 &  0.07 &  0.00 &  0.08 &  0.05 &  0.05 &  0.07 &  0.74 &  1.02 \\
 HD205536  &  5442. &  4.38 &  0.77 & -0.05 & -0.09 & -0.03 & -0.09 & -0.01 & -0.01 &  0.66 &  1.07 \\
 HD206163  &  5519. &  4.43 &  0.94 &  0.01 & -0.07 & -0.03 & -0.18 & -0.05 &  0.01 &  0.51 &  0.93 \\
 HD206172  &  5608. &  4.49 &  0.77 & -0.24 & -0.19 & -0.28 & -0.23 & -0.26 & -0.23 &  0.60 &  1.00 \\
 HD207129  &  5937. &  4.49 &  1.06 &  0.00 & -0.08 & -0.02 & -0.09 & -0.02 & -0.01 &  0.65 &  1.05 \\
 HD207583  &  5534. &  4.46 &  0.99 &  0.01 & -0.58 & -0.03 & -0.21 & -0.05 & -0.02 &  1.55 &  1.00 \\
 HD207700  &  5666. &  4.29 &  0.98 &  0.04 &  0.09 &  0.07 &  0.14 &  0.13 &  0.11 &  0.74 &  1.12 \\
 HD208272  &  5199. &  4.42 &  0.99 & -0.08 & -0.12 & -0.11 &  0.02 & -0.11 & -0.04 &  0.91 &  0.91 \\
 HD208704  &  5826. &  4.38 &  1.04 & -0.09 & -0.04 & -0.11 & -0.09 & -0.10 & -0.08 &  0.59 &  1.02 \\
 HD209742  &  5137. &  4.49 &  0.79 & -0.16 & -0.48 & -0.16 & -0.04 & -0.19 & -0.15 &  1.82 &  0.98 \\
 HD210918  &  5755. &  4.35 &  0.99 & -0.09 & -0.02 & -0.11 & -0.03 & -0.04 & -0.07 &  0.65 &  1.15 \\
 HD211415  &  5850. &  4.39 &  0.99 & -0.21 & -0.22 & -0.23 & -0.20 & -0.17 & -0.19 &  0.69 &  1.12 \\
 HD212580  &  5155. &  4.44 &  0.85 & -0.11 & -0.13 & -0.14 &  0.01 & -0.16 & -0.12 &  0.91 &  0.98 \\
 HD212708  &  5681. &  4.35 &  0.99 &  0.27 &  0.21 &  0.33 &  0.24 &  0.24 &  0.26 &  0.71 &  1.02 \\
 HD213575  &  5671. &  4.18 &  1.02 & -0.15 &  0.05 & -0.12 &  0.01 &  0.05 & -0.03 &  0.60 &  1.29 \\
 HD213628  &  5555. &  4.44 &  0.82 &  0.01 & -0.03 &  0.03 &  0.02 & -0.02 &  0.02 &  0.74 &  0.98 \\
 HD214759  &  5461. &  4.37 &  0.85 &  0.18 &  0.06 &  0.23 &  0.10 &  0.16 &  0.17 &  0.72 &  1.05 \\
 HD215456  &  5789. &  4.10 &  1.19 & -0.09 & -0.20 & -0.11 & -0.08 & -0.06 & -0.06 &  0.87 &  1.07 \\
 HD216777  &  5623. &  4.51 &  0.81 & -0.38 & -0.33 & -0.39 & -0.36 & -0.30 & -0.33 &  0.62 &  1.15 \\
 HD219077  &  5362. &  4.00 &  0.92 & -0.13 & -0.05 & -0.14 & -0.13 & -0.05 & -0.08 &  0.55 &  1.15 \\
 HD219249  &  5482. &  4.50 &  0.74 & -0.40 & -0.44 & -0.41 & -0.34 & -0.34 & -0.36 &  0.83 &  1.12 \\
 HD220256  &  5144. &  4.41 &  0.47 & -0.10 & -0.05 & -0.07 &  0.11 & -0.04 & -0.04 &  0.95 &  1.07 \\
 HD220367  &  6128. &  4.37 &  1.34 & -0.21 & -0.16 & -0.23 & -0.18 & -0.19 & -0.16 &  0.63 &  1.00 \\
 HD220507  &  5698. &  4.29 &  1.01 &  0.01 &  0.12 &  0.03 &  0.11 &  0.11 &  0.05 &  0.65 &  1.23 \\
 HD221146  &  5876. &  4.27 &  1.09 &  0.08 &  0.01 &  0.11 &  0.09 &  0.11 &  0.11 &  0.79 &  1.07 \\
 HD221356  &  6112. &  4.53 &  1.12 & -0.20 & -0.06 & -0.22 & -0.29 & -0.18 & -0.19 &  0.39 &  1.10 \\
 HD221420  &  5847. &  4.03 &  1.28 &  0.33 &  0.21 &  0.42 &  0.26 &  0.37 &  0.36 &  0.74 &  1.10 \\
 HD222335  &  5271. &  4.49 &  0.83 & -0.20 & -0.19 & -0.22 & -0.22 & -0.20 & -0.17 &  0.62 &  1.00 \\
 HD222422  &  5475. &  4.46 &  0.73 & -0.12 & -0.18 & -0.15 & -0.28 & -0.11 & -0.11 &  0.52 &  1.07 \\
 HD222595  &  5648. &  4.46 &  0.88 &  0.01 & -0.05 &  0.02 & -0.03 & -0.01 &  0.02 &  0.69 &  1.00 \\
 HD222669  &  5894. &  4.46 &  1.01 &  0.05 & -0.03 &  0.05 & -0.01 &  0.03 &  0.03 &  0.69 &  1.07 \\
 HD223171  &  5841. &  4.20 &  1.12 &  0.12 &  0.11 &  0.14 &  0.10 &  0.12 &  0.11 &  0.65 &  1.10 \\
 HD223282  &  5328. &  4.49 &  0.60 & -0.41 & -0.36 & -0.44 & -0.31 & -0.38 & -0.37 &  0.74 &  1.05 \\
 HD224393  &  5774. &  4.54 &  0.84 & -0.38 & -0.34 & -0.41 & -0.42 & -0.33 & -0.36 &  0.55 &  1.15 \\
 HD224619  &  5436. &  4.39 &  0.79 & -0.20 & -0.18 & -0.20 & -0.20 & -0.14 & -0.16 &  0.63 &  1.12 \\
 HD224789  &  5185. &  4.44 &  1.05 & -0.03 & -0.01 & -0.03 &  0.06 & -0.07 &  0.01 &  0.78 &  0.89 \\
\hline
\end{tabular}
\end{table}




\clearpage



\end{document}